\documentclass[10pt,journal,compsoc]{IEEEtran}

\ifCLASSOPTIONcompsoc
  \usepackage[nocompress]{cite}
\else
  \usepackage{cite}
\fi

\usepackage{ifpdf}
\usepackage{float}
\usepackage{multirow}
\usepackage{array}
\usepackage{lmodern}
\usepackage{siunitx}
\usepackage{booktabs}
\usepackage{etoolbox}
\usepackage{algorithm, algorithmic}
\usepackage{tabu}

\newcommand{\RNum}[1]{\uppercase\expandafter{\romannumeral #1\relax}}

\ifpdf \usepackage[pdftex]{graphicx} \pdfcompresslevel=9
\else \usepackage[dvips]{graphicx} \fi

\usepackage{t1enc,dfadobe}
\usepackage{xcolor}
\usepackage{color}
\usepackage{subcaption}
\usepackage{graphicx}
\usepackage{amsmath}
\usepackage{flexisym}

\begin{document}

\title{Phoenixmap: An Abstract Approach to Visualize 2D Spatial Distributions}
      
\author{Junhan~Zhao,
        Xiang~Liu,
        Chen~Guo,
        Zhenyu Cheryl~Qian,
        and~Yingjie Victor~Chen 
\IEEEcompsocitemizethanks{\IEEEcompsocthanksitem Junhan Zhao and Yingjie Victor Chen are with the Department
of Computer Graphics Technology, Purdue University, West Lafayette, IN 47907.
E-mail: zhao835@purdue.edu; victorchen@purdue.edu
\IEEEcompsocthanksitem Xiang Liu is with the Department of Computer and Information Technology, Purdue University, West Lafayette, IN 47907.
E-mail: xiang35@purdue.edu
\IEEEcompsocthanksitem Chen Guo is with the School of Media Arts and Design, James Madison University, Harrisonburg, VA 22807.
E-mail:guo4cx@jmu.edu
\IEEEcompsocthanksitem Zhenyu Cheryl Qian is with the Department of Art and Design, Purdue University, West Lafayette, IN 47907.
E-mail: qianz@purdue.edu
\IEEEcompsocthanksitem Dr. Yingjie Victor Chen is the corresponding author.
}
}

\IEEEtitleabstractindextext{%
\begin{abstract}
The multidimensional nature of spatial data poses a challenge for visualization. In this paper, we introduce \textit{Phoenixmap}, a simple abstract visualization method to address the issue of visualizing multiple spatial distributions at once. The \textit{Phoenixmap} approach starts by identifying the enclosed outline of the point collection, then assigns different widths to outline segments according to the segments' corresponding inside regions. Thus, one 2D distribution is represented as an outline with varied thicknesses. \textit{Phoenixmap} is capable of overlaying multiple outlines and comparing them across categories of objects in a 2D space. We chose \textit{heatmap} as a benchmark spatial visualization method and conducted user studies to compare performances among \textit{Phoenixmap}, \textit{heatmap}, and \textit{dot distribution map}. Based on the analysis and participant feedback, we demonstrate that \textit{Phoenixmap} 1) allows users to perceive and compare spatial distribution data efficiently; 2) frees up graphics space with a concise form that can provide visualization design possibilities like overlapping;  and 3) provides a good quantitative perceptual estimating capability given the proper legends. Finally, we discuss several possible applications of \textit{Phoenixmap} and present one visualization of multiple species of birds' active regions in a nature preserve.

\end{abstract}  
\begin{IEEEkeywords}
Data Visualization, Visualization, Algorithms, Geospatial Analysis
\end{IEEEkeywords}}

    \maketitle

\IEEEdisplaynontitleabstractindextext

\IEEEpeerreviewmaketitle
\IEEEraisesectionheading{\section{Introduction}\label{sec:introduction}}
\IEEEPARstart{M}{odern} GIS, GPS, and remote sensing technologies have enabled us to capture large amounts of geospatial data effortlessly. The rich data captured is full of information and can be crucial for analytic practice and decision making. One essential step is to visualize this data semantically on maps for analysts to identify trends and anomalies, or make comparisons. Often, people need to determine how a distribution changes over time, or compare distributions in multiple instances. However, due to rapid increases in amounts of data collected, modern data aggregation strategies face numerous challenges. For example, generic distribution outputs are unable to convey intuitive results to observers and lack specificity about the data's location \cite{Openshaw1984arealUnitProblem}. McPherson and Jetz's research \cite{McPherson2007distribution} attempts to identify broad-scale patterns in the diversity of endangered species' living environments by overlaying their range maps, but finds that the elusive spatial structure in species richness patterns makes it difficult to draw concrete conclusions \cite{McPherson2007distribution}. Practitioners and researchers have developed multiple strategies to visualize data at different granularity levels, from raw data to data in various aggregated forms. Some of these strategies include adding direction, plotting event points on a map, visualizing points as a continuous function as a heatmap, or developing a more elaborate scheme like the Bristle Map \cite{kim2013bristle}. 

Apart from these technical challenges, there remains the unresolved problems of how to visually illustrate the distribution of objects from many different categories and the distribution of one kind of object over time. This is a typical spatiotemporal problem. To demonstrate the distribution of changes over time, one solution is to automatically (or manually) animate a sequence of visualizations in order, as the SemanticPrism system does \cite{Chen2015SemanticPrism}. However, the user can experience cognitive overload if they are required to compare earlier and current scenes, especially if they also need to identify multiple differences among scenes. As a result, it is common to use small multiples of geo-visualizations to compare multiple distributions\cite{Tufte:1986:VDQ:33404}. In doing so, users can compare maps from different cells. However, this conventional solution also suffers from two problems. First, due to the limited screen size, the visualization dimension in each cell may not display sufficient details. Second, it is also cognitively challenging to compare visualizations across multiple cells. In some cases, the "small multiples" approach directly plots points on the map, with colors or shapes to distinguish temporal points. However, this outcome can easily lead to cluttering, which causes perceptional confusions.

\begin{figure*}
   \centering
    \includegraphics[width=.95\linewidth]{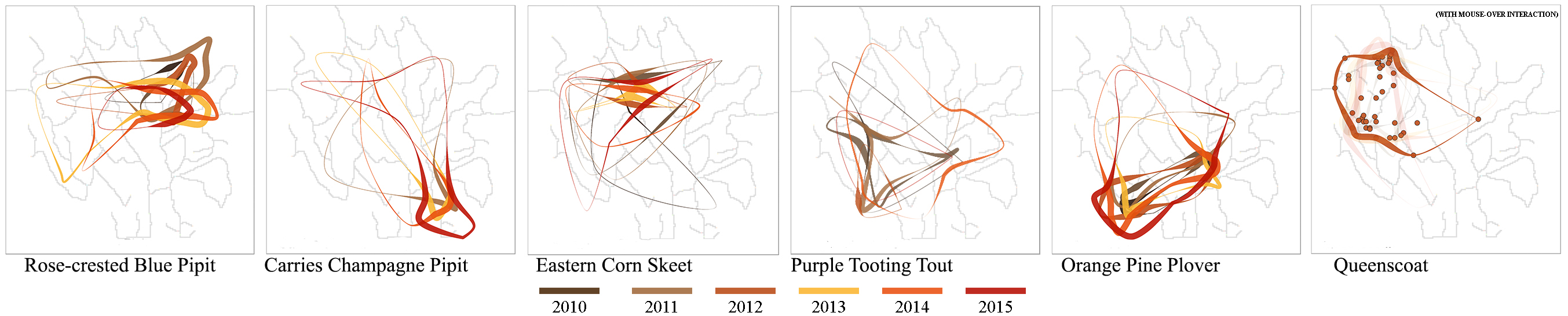}
    \parbox[t]{2\columnwidth}{\relax
    \textit{Phoenixmap} visualizes the patterns of bird distributions and their changes yearly in a national nature preserve (source: 2018 IEEE VAST Challenge~\cite{Phoenixmap}). The width of each segment represents the number of observation records in the area facing that segment. As the colors vary from dark brown to red, the curves show the region changes of birds over recent years. For example, the Orange Pine Plovers' active region has been gradually moving towards the southwest over the years. The graph on the right (Queenscoat) shows that when mousing over a curve in \textit{Phoenixmap}, the observation records will pop out as dots in the area.}
    \caption{\label{fig:teaser} Visualizing birds distribution change overtime in \textit{Phoenixmap}}
\end{figure*}

To address these challenges and deliver sufficient information within the constraints of a thematic map, we introduce and assess an abstract simple spatial distribution visualization--\textit{Phoenixmap} (Fig.~\ref{fig:teaser}). Focused on presenting two fundamental properties, the range and the density of spatial distribution, \textit{Phoenixmap} summarizes and abstracts a spatial distribution into an outline with various widths (thicknesses). The outline can be determined from the boundary of a fixed region (e.g., a room or a political boundary) or computed based on X/Y coordinates using an algorithm such as a concave/convex hull. The width of each segment represents the density of the segment's corresponding area that is adjacent to the segment. Thus, a user can comfortably perceive the region of the objects and quickly estimate the density of subjects in different areas in the region. 

Furthermore, \textit{Phoenixmap} can provide a single view of multiple spatial distributions without interruption. Given its simple visual form, the system can overlap several outlines on the map to represent multiple distributions at once. Assigning a gradient of colors to the curves can show a distribution change over time (Fig.~\ref{fig:teaser}). By assigning colors with different hues to distinguish time periods or species, the user can quickly present multiple distributions for different kinds of objects on the same map. The initial design was developed to solve the 2018 IEEE VAST challenge~\cite{Phoenixmap} and we continued to refine the idea and the algorithm. This paper demonstrates \textit{Phoenixmap} as 1) a robust algorithm, 2) a series of controlled experiments to compare its performance with \textit{heatmap} and  \textit{dot distribution map}, and 3) capable of displaying multiple spatial distributions at once for users to compare or make sense of temporal trends.

\section{Related Work}
\subsection{Spatial Distribution Visualizations}
Spatial distribution data can be widely found in many domains, such as census data, bird migration data, agricultural data, weather forecasts, and water contamination data. Nature boundaries (e.g. coastlines, rivers, roads) or political boundaries often influence these geographical phenomena. The analysis focuses on the outlines of the data points and requires general geographical information to serve as the basis. The difficulty of visualizing spatial distribution data comes from the distribution's widespread scope and variability. Aggregation and parametric statistics are usually employed to visualize the density estimates and map distribution with color, length, height, dot, texture, and glyphs \cite{kao2001visualizing, scott2012multivariate}. 

A box plot is one of the most popular methods to visualize the distribution of a set of values. Tongkumchum's "two-dimensional box plot" \cite{PhattrawanTongkumchum2005Tbp} shows the general characteristics of the 2D points: location, spread, correlation, and skewness. Another variation of the box plot is the "quelplot and relplot" \cite{goldberg1992bivariate}. This uses a pair of ellipses as a hinge and a fence to show location, scale, correlation, and a resistant regression line. Box plot is a simple way to present the spread and skewness of a distribution. However, it cannot display a complicated distribution or provide the exact values and details of the distribution. 

A major challenge in traditional mapping techniques is that any use of dots or colored glyphs for data locations may easily cause overdrawing and hide distribution status. To address this issue, researchers adopt heatmaps to visualize the massive size of the point cloud. Point mapping, choropleth mapping, and kernel density estimation (KDE) are the most widely used heatmapping methods. Maciejewski et al. \cite{maciejewski2010visual} employed KDE heatmaps to estimate event distributions in the spatial realm. These systems also use contour histories to provide the user with a spatiotemporal view of current and past data trends, which is helpful for tracking hotspot movement over time. Chae et al. \cite{chae2014public}  developed a geospatial heatmap to visualize and explore patterns in the spatial distribution of Twitter users over time. This visualization system supports disaster management by identifying abnormal patterns and evaluating varying-density population areas to determine changes in movement. There have been many attempts employing a choropleth map\cite{slocum2008thematic, brewer2002evaluation,meirelles2013design} to encode one variable accurately on a map. A choropleth map displays a colored, shaded, or patterned area in proportion to a data variable. Some research has expanded this idea by adding color blending, color weaving, texture overlays, and animation \cite{hagh2007weaving, panse2006visualization}. Pan et al. \cite{panse2006visualization} combines a cartogram-based layout with a pixel-oriented method to preserve local distributions and avoid overplotting in a large geospatial point set. The pixel-oriented approach trades the absolute and relative positions against clustering for pixel coherence and overlap reduction, but it cannot easily show geographic features and relationships between regions. In contrast, cartograms preserve geospatial positions from the dataset. The combination of cartograms and the pixel-oriented method addresses the overplotting problems while preserving the spatial features of the map. However, these techniques generally occupy a large portion of the screen's real estate. As such, it is difficult to apply these techniques to represent multiple distributions within a single view. 

 To visualize multivariate spatial data, Bristle maps \cite{kim2013bristle} use a variety of graphical properties of a line (length, density, color, transparency, and orientation) to encode a set of variables in large amounts of point data. The topology graph can be used to reduce clutter and overlapping. However, it might be overwhelming for a designer when facing a large number of variables to select a proper combination for visual encoding. Also, the spatial information in the visualization is restricted to "road" level.  To tackle the difficulty of visualizing a multi-dimensional dataset in spatio-temporal distributions, Scheepens et.al \cite{DBLP:journals/tvcg/ScheepensWWAAW11} introduced a method of compositing density maps for aggregating the multivariate trajectories of moving objects.
 
 Scheepens et.al \cite{scheepens2014contour} used contours to visualize traffic-related probability-density fields on spatio-temporal zones. This visualization system is able to predict the future positions of selected moving objects. Tominski et al.\cite{tominski2012stacking} used 2D maps to represent spatial context, and stacked 3D trajectory bands to encode trajectories and attribute data. This visualization tool integrates space, time, and attributes. These visual encoding techniques are particularly effective on traffic-related topics when many objects move along the same routes, but may not be applicable to discern changes of activities spread across a large area.  
 
 While most spatial visualization techniques deal with objects within the screen, several approaches have been developed to show objects outside of the screen \cite{BaudischPatrick2003Hatf,DBLP:journals/corr/JackleFR17}. Outside objects are projected to the boundary area of the display using orthographic or radial projection methods. The Halo \cite{BaudischPatrick2003Hatf} interface used arc with different radii along screen edges to show the direction and distance of objects outside of the screen. Jackle et al.'s method \cite{DBLP:journals/corr/JackleFR17} was able to show numerous off-screen objects in a limited boundary area and maintain the topology relationship among objects. Thus, users can be better aware of relative locations of objects. Although these approaches are designed to show off-screen objects, the projection approach can be applied to on-screen objects. Less important on-screen objects could be projected to a boundary showing contextual information, while keeping the inside screen real estate for more important items. Such an approach could be useful for enhancing the scalibility of the limited screen space. These off-screen approaches focus on dealing with individual objects in space. However, similar to other spatial visualization techniques such as dotmaps, heatmaps and the like, this approach would still not be able to handle data that are large in quantity or change over time.  Our \textit{Phoenixmap} builds upon the concept of projection and is adaptive to different geometric boundaries. 
 
A scatter plot places points on a Cartesian coordinate system to show the relation between two dimensions. One common setback of scatter plots is that overlapping and overplotting can occur when too many objects are plotted, which could potentially lead to occlusion of a significant portion of data values. Keim et al. \cite{doi:10.1057/ivs.2009.34} presented the generalized scatter plot to improve the unsatisfying outcome of a generic scatter plot occluding a significant portion of the dataset shown. This advanced scatter plotter shows a much clearer distribution than traditional scatter plots, but the lack of quantitative estimation caused by overlapping is still a problem. To overcome the overlapping issue, Mayorga \cite{MayorgaA2013SOOi} introduced "Splatterplots" that use contours for bounding the overlapping distributions. However, intersections between the contours when they are combined may cause visual complexity. Furthermore, Splatterplots leave little extra space on the graph when showing the intersections. It limits the graphical space for the designers and hinders its readability.

In the field of spatial distribution visualization, there are many other innovative visualization attempts. Inspired by the effect of concave hull, Li et al. \cite{10.1111:cgf.13414} proposed "ConcaveCubes" to visualize rich information regarding the borders of distributions. With the use of transparency, the graph was capable of showing the base map, but insufficient for presenting the distribution. Neuroth et al. \cite{neuroth2017scalable} used spatially organized histograms in order to distinguish spatial variations and exploit isosurfaces to visualize time-varying trends found within histogram distribution. Wang et al. \cite{wang2017statistical} used warm colors for a high probability of the investigated subjects appearing on the isosurface and cold colors for a low probability. We aim to help users effectively make sense of the spatial distribution of geospatial datasets. The \textit{Phoenixmap} method visualizes the abstraction of the distribution but still preserves spatial information and reduces clutter. In other words, our work focuses on depicting general characteristics of spatial distribution in a more flexible and simplistic manner with greater detail.

\subsection{Outlines and Distributions}
Range and distribution are essential requirements for spatial distribution analysis. Scientists across multiple domains have investigated their subjects' variations, such as health, well-being, and epidemiology, through outline analysis \cite{Jacquez2010}. Our approach is also based on obtaining the outline of a geographic area defined by a given set of points on a map. By retaining the boundary information and abstracting it--projecting the information inside the region to the outline--our solution aims to reduce high-dimensional information to 2D dimensions. 

There exist various algorithms for forming an outline of 2D points. In our demonstration, we selected the concave-convex hull as the primary sample solution. Convex hull and concave hull algorithms establish another foundation for our work representing a set of points distributed on a map. A convex hull of a set of points is defined as the smallest convex polygon that contains all the points in the point set. Sklansky et al. \cite{sklansky1972measuring} proposed an algorithm to compute the convex hull by starting at one point along the convex hull and testing all vertexes for concavity. In contrast, a concave hull contains at least one reflex interior angle. Unlike the convex hull, which is unable to fully reflect the geometric characteristics of a point set, the concave hull is widely used to describe a region occupied by points. Utilizing the K-nearest neighbours algorithm, Moreira's \cite{Moreira2006} algorithm was able to generate convex or concave hulls that can deal with arbitrary sets of points. By choosing a \textit{k} parameter, the user is able to flexibly adjust the shape of the outline to achieve the best result. Later, Wang and Erik \cite{wang2014polygon} utilized a density-based clustering algorithm to distribute polygons into meta-clusters, which the clusters containing sets of similar polygons. Our work relies on a predefined outline, or an outline that is computed from an effective algorithm but not limited or bound to any particular algorithm. In the following implementation, we used the widely accepted concave hull algorithm from Moreira \cite{Moreira2006} for the purposes of demonstration. 

Colin Ware \cite{WareC.2004IVPf} demonstrated a Euler diagram that uses multiple closed contours to represent the overlapping relationships among different data sets. This method is effective because human beings tend to perceive a closed contour as an object (Gestalt laws of closure and continuity). A combination of contour, color, and texture can be used to define the spatial distribution of multiple overlapping regions. In Fig. 1, we used colors as an additional dimension and integrated it with small-multiples to provide more distinguishable visualizations to the end users. Our user test shows that the increased cognition load of this stacking product does not result in negative perceptional difficulties.

\section{Visualization Algorithms}

\begin{figure*}
   \centering
    \includegraphics[width=.9\linewidth]{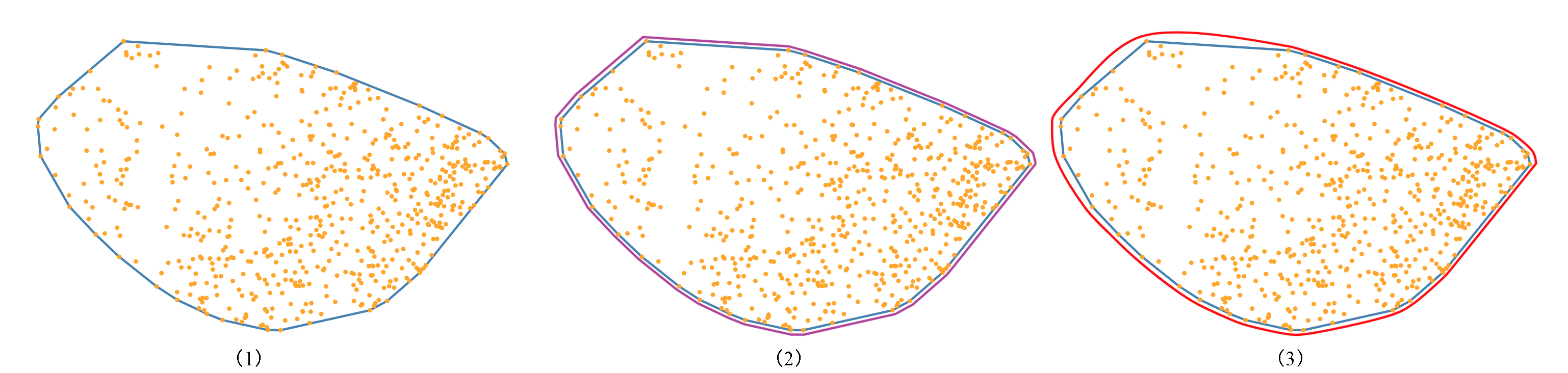}
 \parbox[t]{2\columnwidth}{\relax
(1) Concave hull (blue polygon) (2) Offset the polygon outline (purple polygon) 
  (3) Smooth the polygon to Bezier Curve (red curve)
           }
  \caption{\label{fig:off_bezier} Computing the Bezier Curve Outline }
\end{figure*}

Our proposed method aims to tackle the difficulties in geospatial visualization, including: (1) The cognitive overload caused by the juxtaposition of numerous graphs, (2) The limitations of current geo-visualization methods that do not provide effective solutions to overlapping data points, and (3) The need for more effective use of screen real estate, as in the case of heatmapping; where the visualization leaves little space for displaying other information, because of the large areas covered with colors and shading. Our new approach allows users to perceive and compare multiple geospatial distribution data, where the distribution data belong to various types of objects or the distribution changes over time.

The challenges mentioned above require a simple way of visualizing multiple spatial distributions. In geospatial data, for example, in dealing with distributions of an animal species over time, we normally care about the range and the number of animals in various places within that range. The range can be seen as an outline enclosing many 2D data points. In a heatmap, density is represented by hue and shading, where darker color means higher density; however,  the boundary between high and low densities is usually blurry. In contrast, dot density maps directly reflect raw data and show exact locations, which often makes it difficult for users to estimate density. We also take psychological and aesthetic factors into account. Bertamini et al. \cite{Bertamini2016} conducted a series of experiments on people's perceptions of abstract shapes and discovered that the curved lines were more acceptable than both angular and straight lines for the observers. Hence, we adopt the Bezier curve as one of the bases in our algorithm.  

Here, we introduce a new approach to represent both range and density. For a group of points on a 2D map, first we define an enclosed curve as the outline, then we assign different widths to segments of the outline to show the point density in the areas within and adjacent to each curve segment. Our approach helps users to  estimate the quantitative density and distribution of the data more quickly when they use \textit{Phoenixmap} for data interpretation. 

\subsection{Outline Computation}

With respect to data aggregation, our method first requires an outline to surround the data points. In a real-world problem, there are two ways of determining such an outline: 

\begin{itemize}
    \item Use a predefined outline. Sometimes objects are distributed within a fixed scope. For example, consider observing objects' movements in a room. The walls make a natural outline regarding the range of the spatial distributions. For geographic regions, outlines can  be derived directly from political boundaries or natural topography like coastal lines and rivers.   
    \item Compute an outline. When objects are distributed freely in nature without a natural physical or political boundary, we may need to compute the outline from the data points. 
\end{itemize}

There are numerous modern methods capable of computing an outline, as we reviewed previously. For demonstration purposes, we selected a "concave hull" method for computing the edges\cite{Moreira2006}, due to its flexibility and the simplicity of the final result. Any other reasonable outline computation algorithm (e.g.,\cite{wang2014polygon}) could be applied. Utilizing different clustering and outline algorithms, we can provide a single secure border for the whole population or multiple tight outlines for sub-groups.

\begin{algorithm}[H]
\caption{\label{algor1}\textit{Phoenixmap}}
\begin{algorithmic}[1]
\REQUIRE A set of points $P=\{p_1,p_2,\dots,p_k\}$ where $p_i$ is the coordinates for $i=1, \dots, k$; number of segments $n$
\ENSURE \textrm{\textit{Phoenixmap}}
    \STATE $B \gets$ Compute outline based on $P$ (e.g. user defined or using an algorithm e.g. concave hull)
    \STATE $B^\prime \gets$ $B + b$ where $b$ is a small positive number. $B^\prime$ is the offset outline. See Fig~\ref{fig:4in1}(3).
    \STATE Use $B^\prime$ to generate a \underline{closed} Bezier curve $C$ using a curve fitting algorithm.
    \STATE Divide $C$ into $n$ segments to get $V$, a set of divisor points and $S$, a set of segments. 
    \STATE $W \gets \emptyset$
    \FOR{$i \gets 0$ to  $n-1$}
        \STATE $v_i \gets$ a point from $V$
        \STATE $v_{i+1} \gets$ adjacent point of $v_i$. Note, $v_n$ is $v_0$.
        \STATE $c_i \gets$ the inscribed circle inside $C$ with respect to $v_i$
        \STATE $o_i \gets$ the center of the circle $c_i$
        \STATE $r_i \gets$ the radius of the circle.
        
         \STATE Define a polygon region $R_i$ through the points $ ( v_i, v_{i+1}, o_i, o_{i+1} )$.
        \STATE $Sum_i \gets$ the number of points that belong to $P$ inside the region $R_i$, and $A_i \gets$ the area of $R_i$.
        \STATE $d_i  \gets Sum_i /  A_i$ ( $d_i$ is the density).
        \STATE Append $d_i$ to $W$
    \ENDFOR
    \STATE Set the sliding window as size 2$x$ and apply \underline{weighted arithmetic mean} (WAM) calculation on all density outputs, using the area $A_i$ of every individual slice as weight factor for WAM. ($x$ is suggested to be a divisor of $n$ as an \underline{empirical number}, e.g. if $n$ is set as 3000, $x$ should be 75, 150, 300, or 500) 
    \FOR{$i \gets 1$ to  $n$}
        \STATE $w_i^\prime  \gets \sum_{k=i-x}^{i+x}(w_k \times A_k)/\sum_{k=i-x}^{i+x} A_k$, $k \in [i-x, i+x)$
        \STATE $w_i^\prime$ = $w_i^\prime \times c \gets$ constant scalar to scale the number up or down. Assign $w_i^\prime$ to the stroke width of the segment $s_i \in S$. 
    \ENDFOR
    \STATE Connect all segments to be the curve $C^\prime$
    \STATE Clip out the outer half of $C^\prime$ using the original Bezier $C$
    \STATE \textbf{return} $C^\prime$ as the \underline{\textit{Phoenixmap}} for $P$
\end{algorithmic}
\end{algorithm}

Using concave hull and taking a set of geospatial data,  $P$, as the input, our \textit{Phoenixmap} algorithm initially generates a polygon outline denoted as $B$. Adding bias $b$ to $B$ pixel-wise, we offset the border by the bias $b$ towards the outside to get $B^\prime$. The purpose of offset is to make sure all points will be included in $B^\prime$. Also, $B^\prime$ yields space for edge thickness computation in the following steps. We extract the control points set $P^\prime$ and compute a closed Bezier curve without self-intersection to go through $P^\prime$. One reason we choose a Bezier curve is that we can make the curve smooth with  $C_2$ continuity and still have it pass through all the control points. According to design principles discussed by Bertamini et al. \cite{Bertamini2016}, people prefer smooth curvature over angularity. This curve is a closed curve that does not intersect itself, since the concave hull is not self-intersecting \cite{Bezier2009}. We denote the outline as $C$. See Fig.~\ref{fig:off_bezier} for this computing process.

\subsection{Segmentation and Thickness Computation}
We intended to use width (thickness) to represent the density of points enclosed by the region. However, in a distribution, density varies in different regions. Thus, we decided to break the outline into many segments, giving each segment a different thickness to represent the density enclosed by its corresponding inside region. For each segment, a user should easily see the density of the region represented by the segment without confusion. Here we use the following method to achieve this goal.  

Aiming to measure the distribution, we differentiated the outline $C$ into $n$ segments. The divisor points are denoted as set $V$, and the segments are denoted as $S$. In our experiment, 3000 is the empirical number of segments considering the number of points, computing power, and effectiveness. A larger segmenting number like 3000 can divide the original curve into small enough segments so that each fits in one pixel. The idea is similar to differential geometry.

\begin{figure}[H]
   \centering
    \includegraphics[width=.9\linewidth]{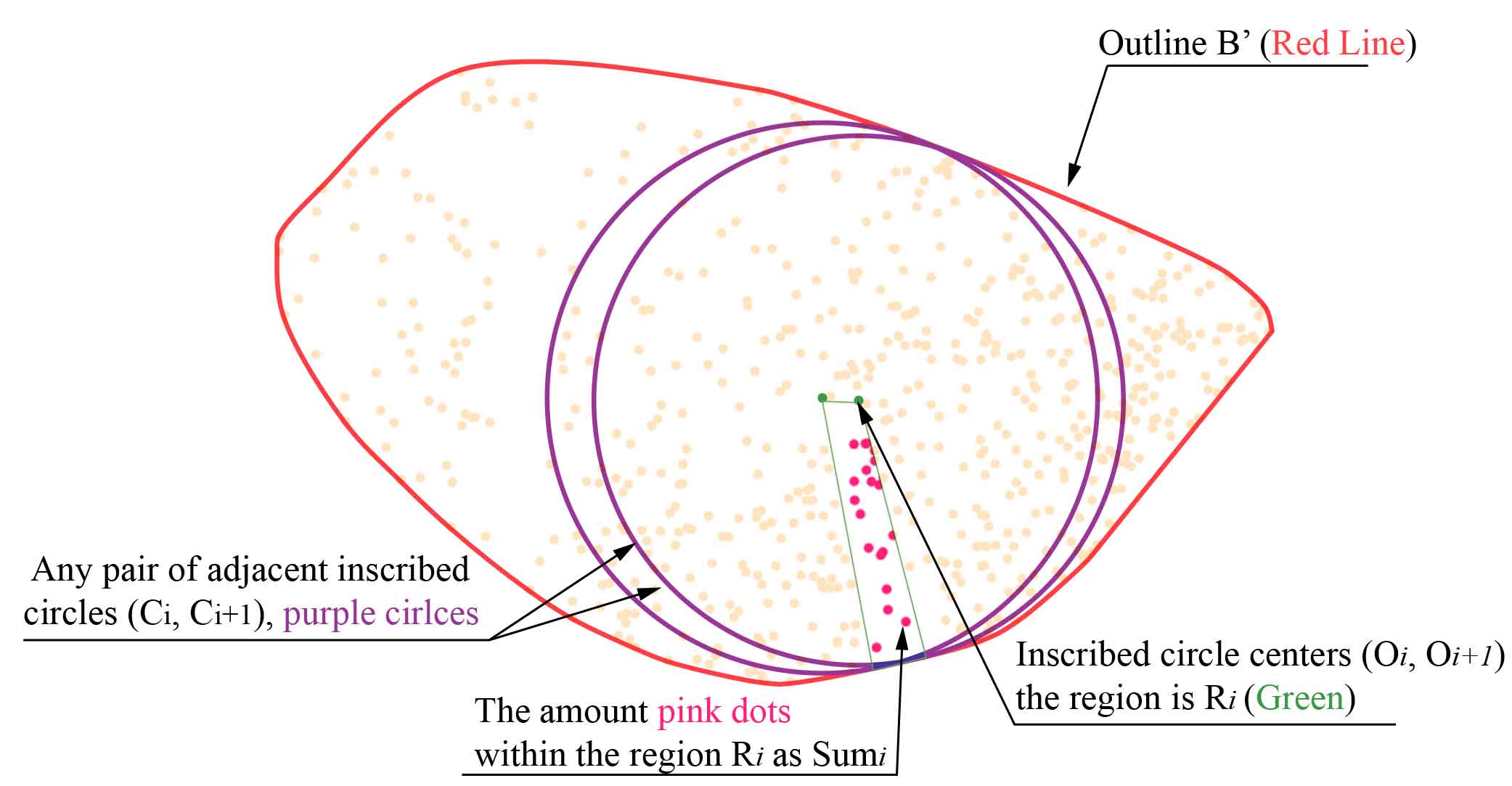}
  \parbox[t]{.9\columnwidth}{\relax
 The two purple curves are inscribed circles with respect to two adjacent points. The green polygon is the region, and the pink points show the selected points within that region.
           }
  \caption{\label{fig:inscribe} Define the Region by Computing the Inscribed Circle.}
\end{figure}

Our ultimate goal is to find a simple way to represent varying densities inside the outline. Our design uses the width (thickness) of the outline to represent the density of points contained in subsections of the overall range. Since most of the outlines of the distributions are irregular, dividing the outlines into small pieces increases the variety and the accuracy of the computed areas. For any segment $s_i \in S$, we defined a region to count the number of subjects. To gain the region, we first compute the inscribed circle through one end point $v_i$, and mark this circle as $c_i$, where $o_i$ and $r_i$ are its center and radius respectively. For each pair of adjacent points ($v_i$, $v_{i+1}$), we construct a region $R_i$ from $(v_i,v_{i+1}, o_i,o_{i+1})$. $R_i$ and $R_{i+1}$ will be next to each other. It is provable that the slices together cover the entire area without intersecting with each other. In Fig.~\ref{fig:inscribe}, one can see the two adjacent inscribed circles. The quadrangle composed by the pair of circle centers and radius end points is highlighted by a green line. The pink dots within represent the enclosed objects. We count number of objects inside each region and denote it as $Sum_i$.

\begin{figure}[!htbp]
   \centering
    \includegraphics[width=.8\linewidth]{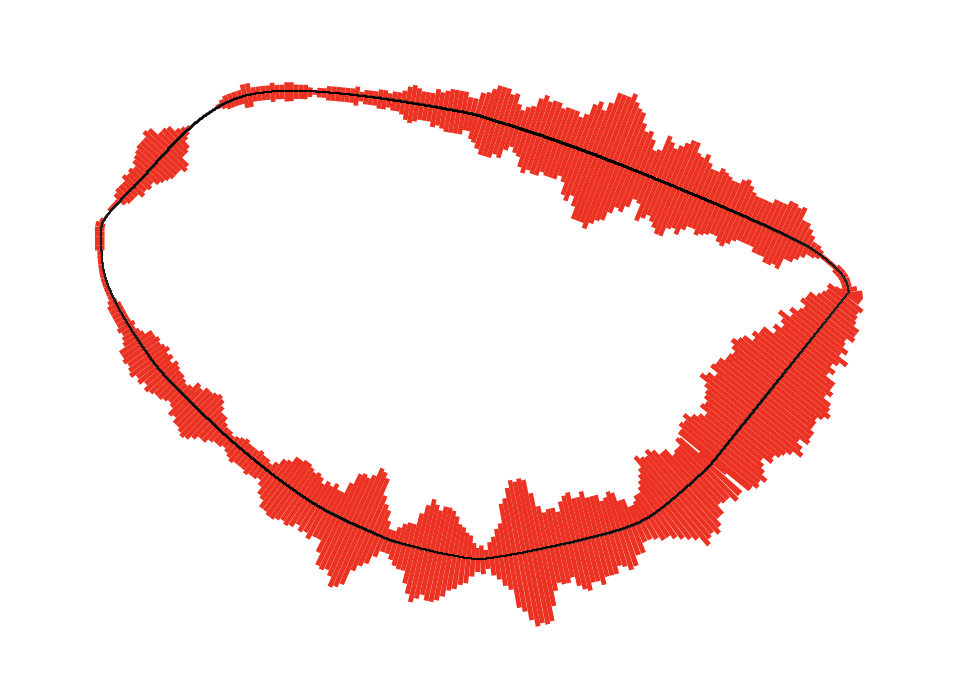}
  \caption{\label{fig:discrete} The discrete segments sit along the curve. This outline is divided into 3000 segments.  The width of each curve segment is computed from the density of corresponding areas inside.}
\end{figure}

 Dividing $Sum_i$ by the area of slice $A_i$, we get the density $d_i$ of that individual slice (see the pink dots in Fig.\ref{fig:inscribe}). Repeating this for each slice to generate a list of densities, we then appended each density $d_i$ to the width ($W$) of each corresponding segment. See Fig.~\ref{fig:discrete} for the discrete segmented curve. With the large number of short segments along the curve, the width (thickness) of the segments varies greatly, since distribution will not be uniform in the real world. For an overview abstract visualization, the fine details in Fig.~\ref{fig:discrete} are not necessary, nor do they look very visually appealing. In the next step, we smooth these segments into a continuous curve.
 
\subsection{Discrete to Continuous: Segment Smoothing}

 To convert the discrete graph to a continuous smooth curve, we apply a sliding window mechanism to smooth the zigzag segments. This involves computing a smoothed, weighted width $w_i^\textprime$ for $w_i \in W$. We introduce another empirical number for the size of the window, denoted as $x$. Based on our experiments with different values, we suggest that $x$ should equal between one-sixth and one-twentieth of the total number of segments $n$. For instance, if $n$ is 3000, then 150, 300, or 500 would be preferred values for $x$. If $x$ is too big, e.g. more than one-tenth of $n$, the curve will be very smooth but may lose detail. If $x$ is small, e.g. less than one-twentieth of $n$, the curve carries a lot of detail but may look rugged. When computing, each segment carries a weight of its corresponding area $A_i$. Some segments correspond to large areas, while others correspond to small areas. Given any $w_i \in W$, we compute the sum of the weighted width $w_k \times A_k$ within 2$x$ size from $w_{i-x}$ to $w_{i+x}$. In the end, we compute the weighted average of $w_i^\prime$ and replace $w_i$ in the width list.

 After this transformation, the width (thickness) of neighbouring segments is averaged out to convert the discrete weighted segment curve into a continuous weighted curve. After the sliding window treatment, we denote the new smooth curve as $C^\prime$.
 
\begin{figure*}
   \centering
    \includegraphics[width=.95\linewidth]{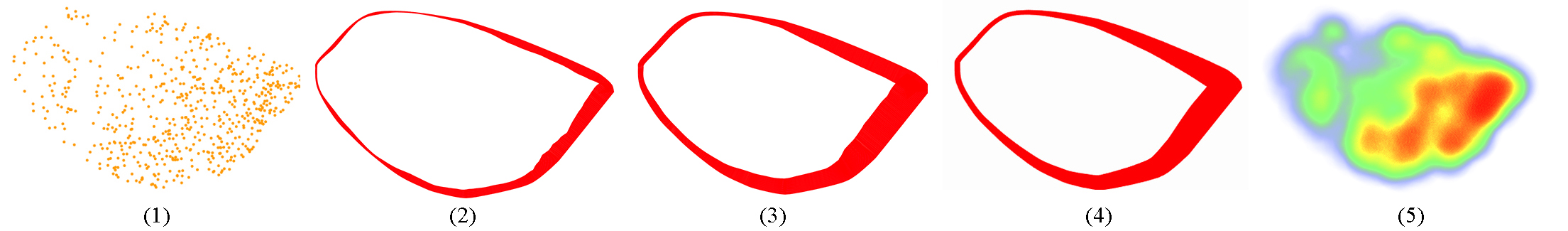}
    \parbox[t]{2\columnwidth}{\relax
(1) Original Data (2) Segments = 3000, Sliding Window = 70
  (3) Segments = 3000, Sliding Window = 150
  (4) Segments = 3000, Sliding Window = 300
  (5) Heatmap
           }
  \caption{\label{fig:4in1} Visualizations using different combinations of empirical numbers. The larger the number of segments and the larger the sliding window, the smoother the curve. A large sliding window may make the curve lose details.}
\end{figure*}

Due to its width, the outside edge of $C^\prime$ will extend beyond the original computed Bezier curve outline. To make the calculated curve accurately represent the original outline, we use the original Bezier curve outline $C$ to clip the $C^\prime$ and only keep the inside half. The graph then appears similar to a plot of consecutive density bars along non-linear coordinates, which is the curve $C^\prime$. Fig.~\ref{fig:4in1} shows four versions of the final continuous curve, demonstrating the various visual effects given by using different empirical numbers (segment numbers and sliding window sizes) and combinations. Given a larger sliding window, (4) is much smoother than (2), but has lost small details.  

Now the thicknesses of the outline corresponds to the densities of the sub-regions bounded by the outline. For a distribution, assuming the objects are densely located in a smaller area, there will be a short wide segment appeared in the outline due to the higher density inside. On the contrary, if the objects are sparse in a  area, the corresponding part of the outline will be thin.

\subsection{Legend Design}

\begin{figure}[H]
   \centering
    \includegraphics[width=.95\linewidth]{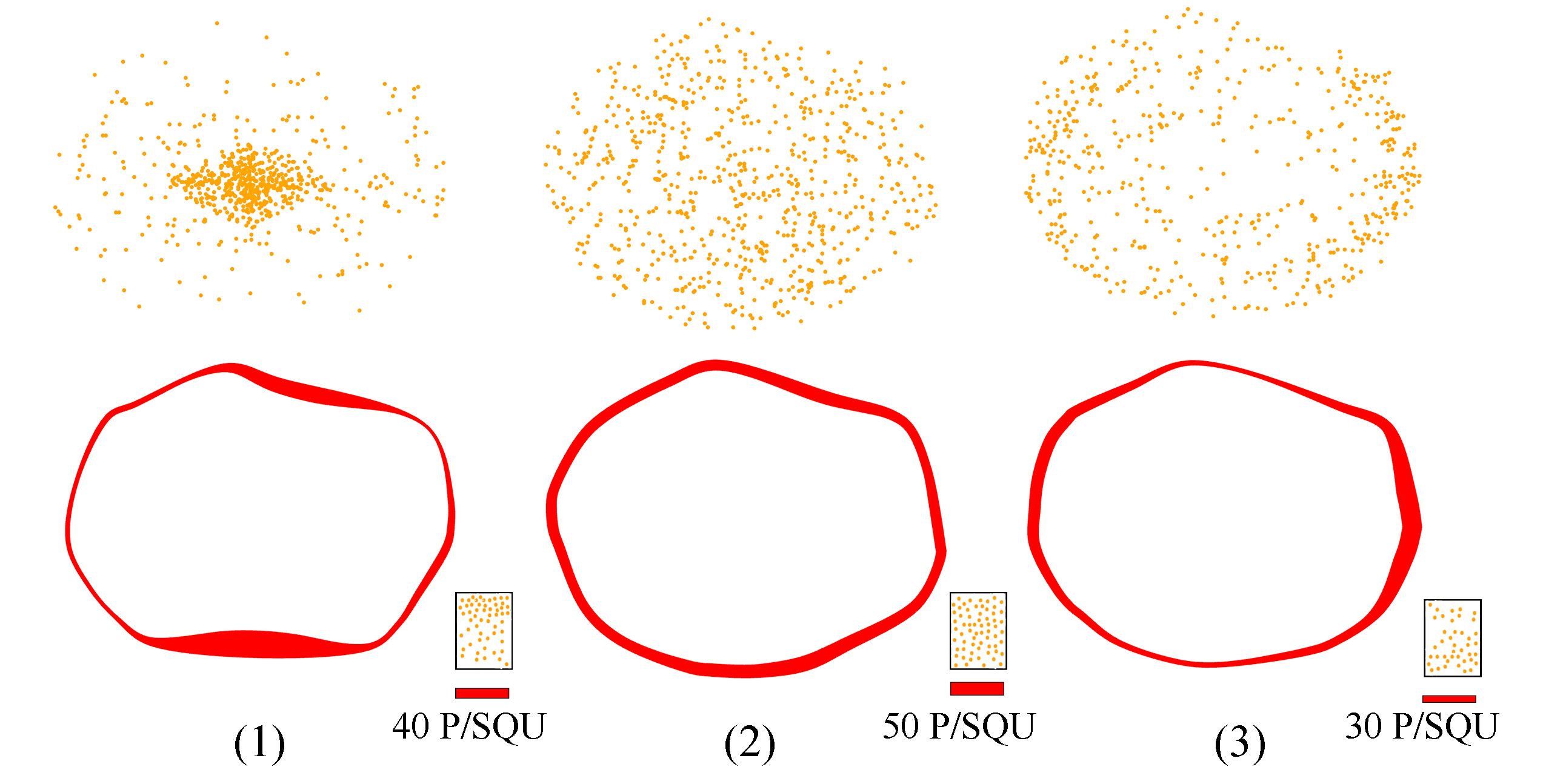}
  \parbox[t]{.9\columnwidth}{\relax
The assistant legends show the prime density distribution from the central region to the outline. The box-legend shows the distribution in a unit. The bar-legend with the text shows the points in the box, which is also the corresponding region's thickness on \textit{\textit{Phoenixmap}}. P means points, and SQU means square units. 
           }
  \caption{\label{fig:legend} Legends for Density and Different Types of Distributions.}
\end{figure}

The algorithm precisely visualizes the outline and density information for the distribution, but one can notice that within the minimal sliced region, the visualization omits the distribution details. To provide more information, we designed a legend to further represent the major distribution characteristics inside the region and to help on quantitative estimation. Fig.~\ref{fig:legend} demonstrates the generic distribution trends for three different scenarios: normal distribution, in which objects are more dense toward the center of the region; uniform distribution, in which objects are scattered uniformly in the range; and close-to-edge distribution, in which objects are more dense toward the edge. The users can quickly view the supplementary legend box and see the estimated gradient density distribution inside the outline. In each legend, the top of the box represents the density close to the approximate geometric center, while the bottom shows the density close to the edge. 

To compute the density legend, for each segment slice in Fig.~\ref{fig:inscribe}, from the center to the edge, we divide the slice into many (e.g. 100, as in the example) sub-sections. We then compute the average density of each sub-section from all slices. Thus we can get an average distribution. We map the average distribution into a box and also introduce a short line to quantify the thickness of the \textit{\textit{Phoenixmap}}'s outline. With the text at the bottom of the legend, the user can easily see how many points correspond to the specific thicknesses shown in the graph, and understand the characteristic of the distribution. In the text description, P stands for points and SQU is square units. The unit should be defined based on the applicaton, e.g. in inches, feet, meters, or kilometers. In Fig.\ref{fig:legend}, we only put one bar legend as a sample. In a real case, the developer can put a series of  legends with different thickness to help the reader better estimate densities (see legends in Fig. \ref{fig:quant}). 

This legend helps the user to gain a deeper understanding about the formation of \textit{\textit{Phoenixmap}}. In the experiments described in section 4, we utilized multiple legends with various widths to guide users' estimations.

\section{Experiments}
To evaluate this new visualization method for spatial data distribution, we conducted three human subject experiments to examine the efficiency and efficacy of information delivery by \textit{\textit{Phoenixmap}}. We compared \textit{Phoneixmap} with \textit{heatmap} and \textit{dot distribution map} (dotmap). In our experiment, all heatmaps use a blue-red color scale, since this color scale is  widely accepted  and  considered as a default color for most heatmap generators (ggplot2 in R, heatmap.js in Javascript, and cyclic color in matplotlib for Python)~\cite{ggplot2, hunter2007, heatmapjs}. Also a multi-color scale has been proven better than single hue in heatmaps~\cite{rogowitz1998data,ware1988color,liu2018somewhere}.  College students from STEM disciplines as well as arts and design fields were recruited for the evaluation tests. For the first two experiments, there were 52 participants, 60 percent male and 40 percent female. We built an online survey website to support user input and time recording. 

 In the first experiment, we selected and simulated typical distribution scenarios and asked users to quickly match the given dot distribution graphs to the \textit{Phoenixmap} or heatmaps. The ratio of the correct matches will suggest the successful recognition of the distribution density and outline.

In the second experiment, we randomly generated spatial distributions with more than 100 points. The users were asked to rank selected areas in the visualizations (either dots, heatmap, or \textit{Phoenixmap}) from most to least dense. We recorded the time taken on each individual task and calculated the correctness ratio. 

The third experiment tested how well participants perceived and quantitatively estimated the sample population density using a \textit{Phoenixmap} compared to when using a heatmap. We recruited 64 participants in total and divided them into two groups. One group took the test with heatmaps while the other used \textit{Phoenixmaps}.

For all the tasks, we asked the participants to perform as quickly as possible while maintaining a comfortable level of confidence about their accuracy. We recorded the time taken and error rate for each participant.

\subsection{Experiment \RNum{1} Design}
For the first experiment, we devised three typical scenarios for testing, each corresponding to a possible real-world scenario. There were two tasks, Task 1 and Task 2, each containing subtasks depending on which dataset was assigned. In Task 1, Subtask A contained eight heatmaps, and Subtask B eight \textit{Phoenixmaps}. In Task 2, Subtask C had 16 heatmaps, and Subtask D had 16 \textit{Phoenixmaps}. Participants were alternately assigned to a starting subtask (A to D) and followed the task sequences to complete all four subtasks. This guaranteed the randomization of the experiment. 

\subsubsection{Training: Regular Geometric Outlines}
\begin{figure}[!htbp]
   \centering
    \includegraphics[width=.9\linewidth]{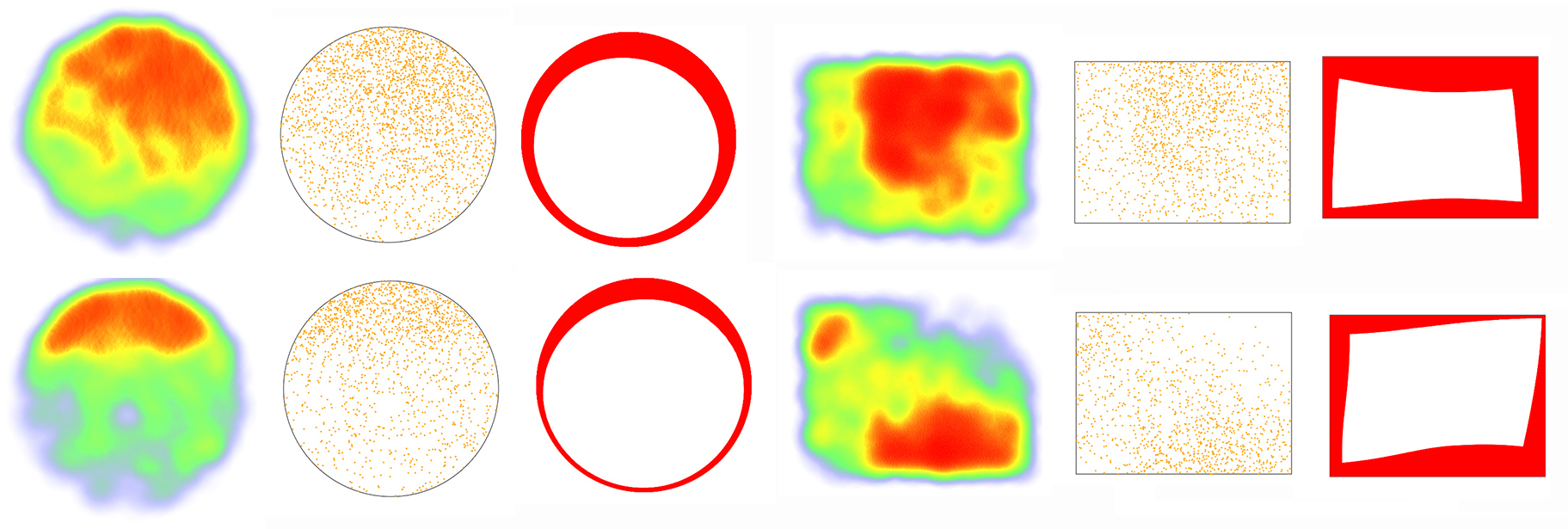}
  \parbox[t]{.9\columnwidth}{\relax
    Two distributions visualized with heatmap, dotmap, and \textit{Phoenixmap}. One was bounded by a rectangle and the other was by a circle.
           }
  \caption{\label{fig:rectcircle} Training Task (Selected Graphs)}
\end{figure}

In this scenario, one can imagine the investigated objects (animal or atom movement) as inside a enclosed space (a room or preserve). The spatial distribution changes inside a fixed outline. We selected the rectangle and circle as basic outlines and arbitrarily generated points inside these outlines. For each outline, we generated eight distributions with different densities across the region. This task served as a training process, an important step considering that \textit{Phoenixmap} visualizations would be new to the users, while the heatmap is more familiar. For fair comparison, the same sizes of circles and rectangles have been used in each graph. Fig.~\ref{fig:rectcircle} demonstrates two of the eight sets of visualizations from the training task. Each of the eight sets of distribution data was generated in all three graph forms, so there were 24 distinct visualizations total.

During the training, participants were briefed about the purpose of the study, about what heatmaps and \textit{Phoenixmaps} were, and about what kinds of tasks they would be asked to perform. The participants then completed the eight training tasks. There were three columns of the visualizations in the task: one heatmap, one dotmap, and one \textit{Phoenixmap} from left to right. Within each column, the graphs were presented in a randomized order, and the participants were asked to match the circle or rectangle dotmaps in the middle with the relevant \textit{Phoenixmap} on the right and the heatmap on the left showing the same spatial distribution and density (Fig.~\ref{fig:rectcircle}). Each row contained three random distribution visualizations. This training will help the participants understand how to map segments on the outline to the density of internal areas.  Before proceeding to the next stage, each participant was given the correct answers for each training task.

\subsubsection{Task 1: Irregular Geometric Outlines}

\begin{figure}[H]
   \centering
    \includegraphics[width=.7\linewidth]{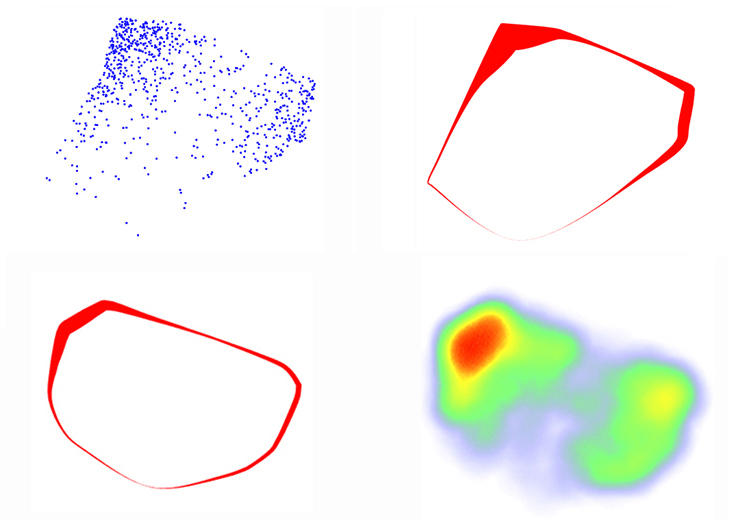}
  \parbox[t]{.7\columnwidth}{\relax
The participants matched a dot distribution to a heatmap or a \textit{Phoenixmap}.}
  \caption{\label{fig:task1} Task 1: Outline Variation (Matching boundaries)}
\end{figure}

We generated 16 distributions with different outlines and densities. Eight of them were converted into heatmaps for Subtask A, while the other eight used in Subtask B for \textit{Phoenixmaps} with outlines generated using the concave hull algorithm \cite{Moreira2006}. We randomized the order of the heatmaps and \textit{Phoenixmap}s. The users were asked to match a dotmap to the corresponding heatmap or \textit{Phoenixmap}.  Fig.~\ref{fig:task1} shows a example of the matching task. The user need to match the dotmap to a heatmap or a \textit{Phoenixmap}. 

\subsubsection{Task 2: Same Outline with Various Distributions}

\begin{figure}[H]
   \centering
    \includegraphics[width=.9\linewidth]{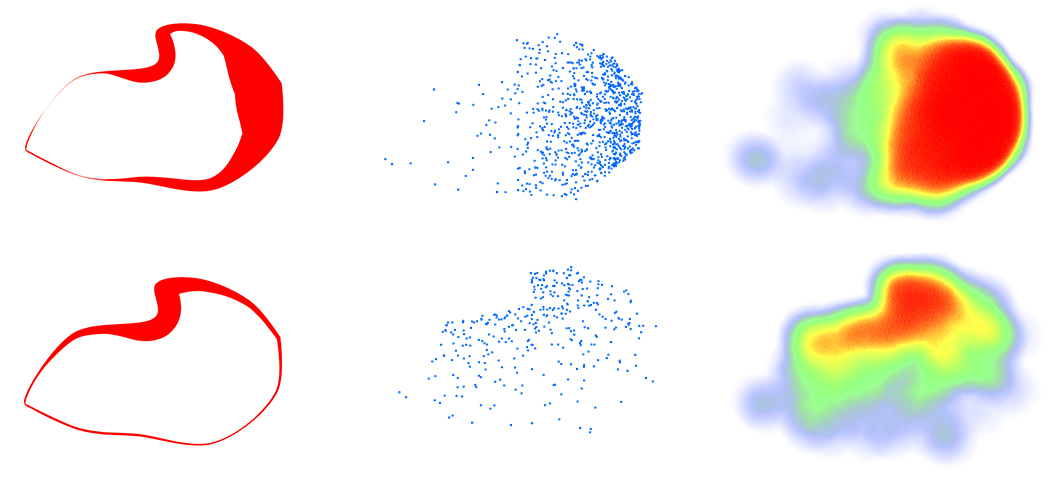}
  \parbox[t]{.9\columnwidth}{\relax
In Subtasks C and D examples, the outline is similar but with different densities in distributions. A participant needs to match a dotmap to a heatmap or a \textit{Phoneixmap}.  }
  \caption{\label{fig:task2} Task 2: Distribution Variation (Matching densities within similar boundaries)}
\end{figure}

In a real-world scenario, it is possible that different distributions fall within the same outlines. Without the outline difference as a basis for comparison, the difficulty of distinguishing the distributions will increase. To simulate this same outline situation, we randomly generated four different outlines with four distributions for each outline, total 16 heatmaps and 16 \textit{Phoneixmaps}. We separated the corresponding heatmaps and \textit{Phoenixmaps} into sub-task C and sub-task D. We asked all 52 participants to perform both two tasks.  Fig.~\ref{fig:task2} shows a example.

\subsection{Experiment \RNum{2} Design}

We designed Experiment \RNum{2} to test whether users can distinguish densities at different locations within a larger distribution. We randomly generated three spatial distributions. Each of the distribution maps contained more than 100 points so as to avoid sparse distribution (for which a simple dot density map is sufficient for visualizing, since all the subjects are able to be counted in the graph independently). 

\begin{figure}[H]
   \centering
    \includegraphics[width=.95\linewidth]{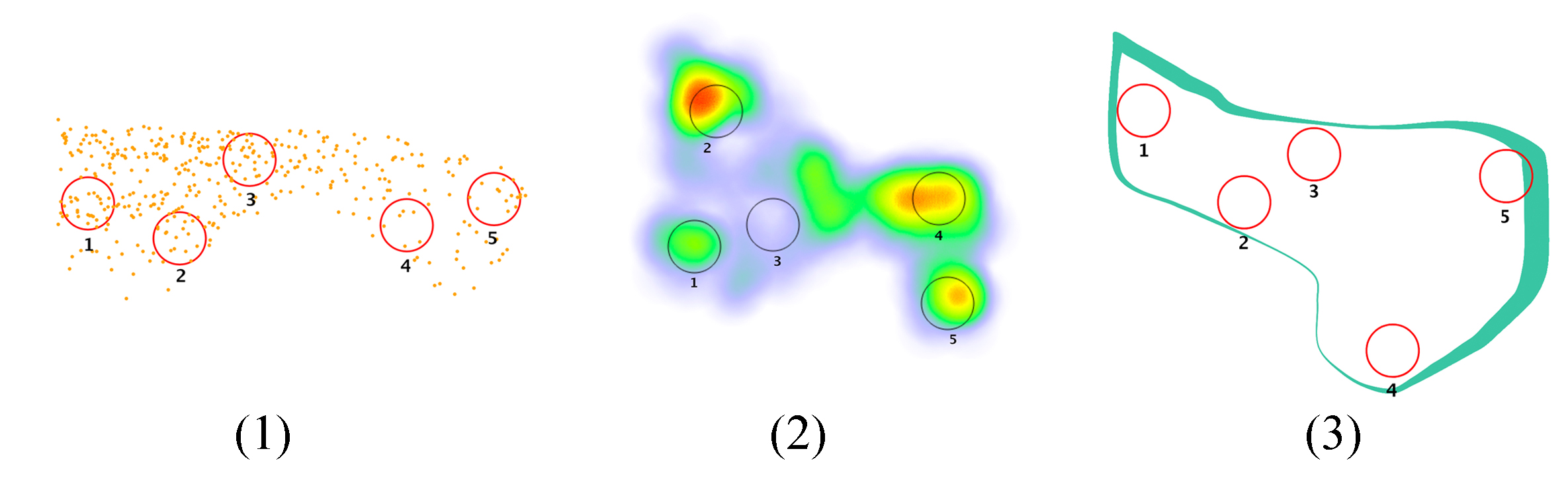}
  \parbox[t]{.9\columnwidth}{\relax
Each participant ranked the relative densities of all five circled areas within the three distribution visualizations. 
           }
  \caption{\label{fig:exp2} Ranking Task: Densities within Distributions}
\end{figure}

Within each distribution, we randomly selected five smaller areas, which were marked out by circles with same radius (Fig.~\ref{fig:exp2}). Therefore, the density of each circle can be regarded as the number of subjects in the circle. We purposely made all five unique densities. These circles were large enough to cover many points to avoid random sample errors.

Each participant was asked to complete density ranking for all three visualization types. However, participants only saw one type of visualization for each distribution. For instance, one participant received a task like Fig.~\ref{fig:exp2}, where three distributions was represented in three different visualization methods respectively. The three distribution maps displayed in our system in a random sequence, so that the system could record the time taken and the user-typed ranking answers. We asked the users to order the rank from "dense" to "sparse." All the participants from Experiment \RNum{1} participated in Experiment \RNum{2}, because Experiment \RNum{2} required a deeper understanding about the generating algorithm of \textit{Phoenixmap}. 

\subsection{Experiment \RNum{3} Design}

The goal of this experiment was to evaluate users' capabilities for perceptual estimation of densities using the \textit{Phoenixmap}s with legends introduced. We conducted user tests on three different distributions, similarly to the last experiment.  We recruited 64 participants. Half of these had participated in the previous two experiments, while the other half were totally fresh to the experiment. Participants received training and introduction on both \textit{Phoenixmap} and heatmap. Each participant was randomly assigned to three visualizations mixed with \textit{Phoenixmap} and heamap, but distinct in distribution. Our online testing system made sure each distribution was tested evenly. 

\begin{figure}[H]
   \centering
    \includegraphics[width=.95\linewidth]{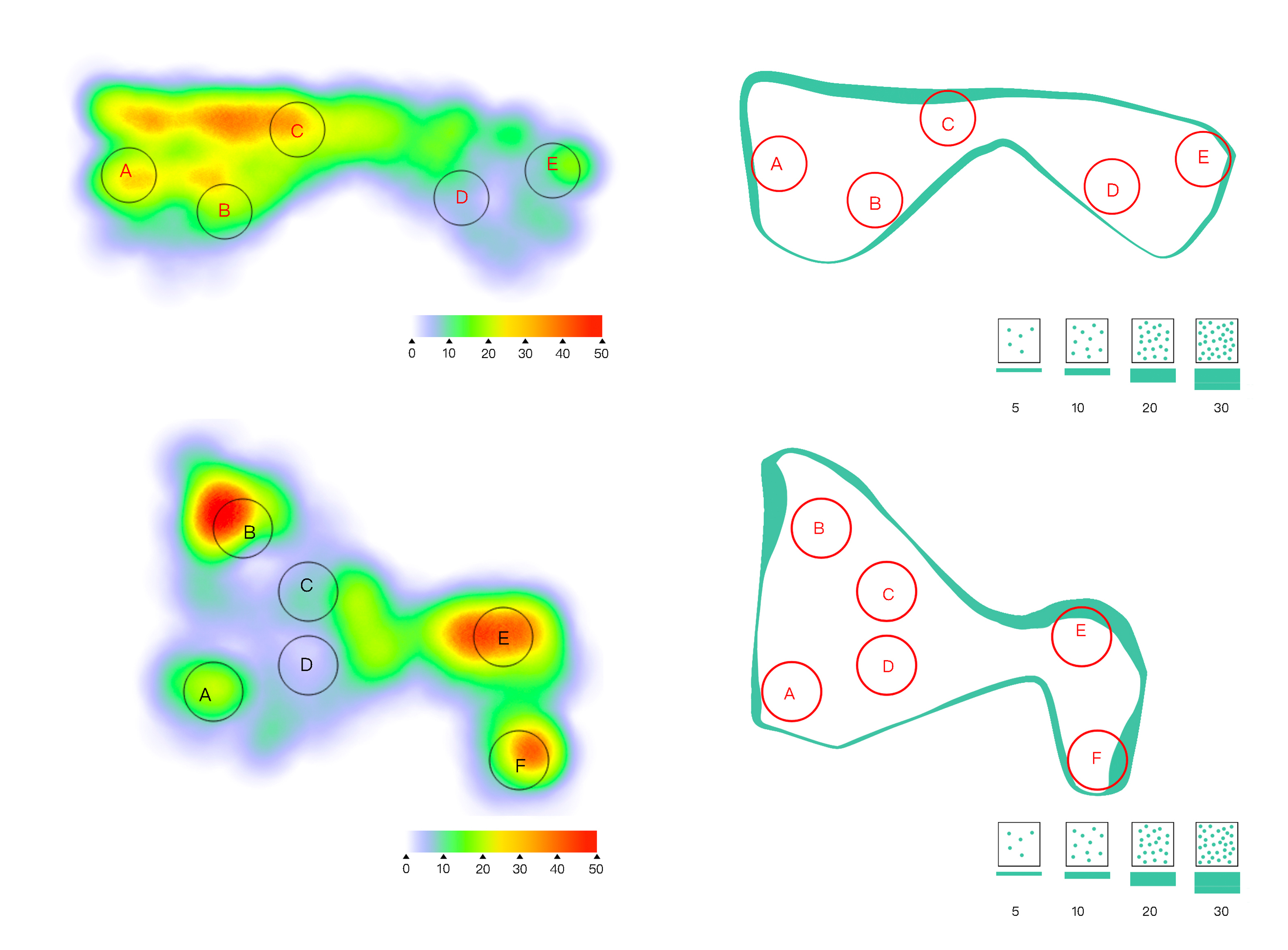}
  \parbox[t]{.9\columnwidth}{\relax
As part of this task, participants may receive distributions visualized by either a \textit{Phoenixmap} or a heatmap with legends associated. Two examples are shown in this figure.            }
  \caption{\label{fig:quant} Quantitative Perceptual Estimation Task}
\end{figure}

Fig.~\ref{fig:quant} shows an example of this experiment. Users were asked to estimate the densities of the circled areas. We located some circles in the center to increase the difficulty of the estimations, since for such areas users would need to reference multiple sides of the outline to estimate the density (See circle D in the upper set of graphs or B in the bottom set from Fig.~\ref{fig:quant}). Notably, many of the circles contain a combination of colors or correspond to an outline segment with varying width. The participants can not simply choose the maximum value. 

\subsection{Qualitative Interview and User Feedback Survey}

At the completion of all experiments, we invited these participants to answer our survey regarding their attitudes toward \textit{Phoenixmap}. The questions covered their demographic information, the strategies they used to solve \textit{Phoenixmap}, and their evaluations of \textit{Phoenixmap} on comprehension and effectiveness. We also interviewed the participants about their preferences and their reasons for these. 

We selected a dataset from IEEE VAST Challenge 2018, visualized it in multiple ways for our participants (See Fig.~\ref{fig:sixcolor}), and then asked whether they could see the distribution of the six species of birds. On the dotmap this data is messy, considering the dots (each representing one bird record) are overlapping. A user may notice some very dense areas, but will have a hard time seeing the full activity range of each bird. 

Additionally, we invited the participants to express their expectations and comments regarding spatial distribution visualization and analysis, more specifically asking whether they would consider using \textit{Phoenixmap} for the visualization of spatial data in their own work.

\section{Results}

After testing the normality of the all the data collected, we determined that the distribution of results was not clearly normal. Without assuming the results as a normal distribution, we applied a paired non-parametric statistical method, the Wilcoxon signed-rank test \cite{CrichtonN2000Wsrt}, to the results. This method can be nearly as efficient as the t-test is on normal distributions.

\subsection{Experiment \RNum{1}}

We evaluated the accuracy of participants' matching work and their finishing times when using either \textit{Phoenixmap} or heatmap. To measure participants' accuracy, we computed the percentage of correct answers. 

Table~\ref{tab:Stas} illustrates that during Experiment I, Task 1 (see Fig. \ref{fig:task1}), users finished on average 10 seconds faster when using heatmap graphs than when using \textit{Phoenixmap}. The average time cost was 54.46 seconds for heatmap and 64.06 seconds for \textit{Phoenixmap}. In terms of accuracy, \textit{Phoenixmap}'s performance was slightly better than that of heatmap (96\% versus 94\%). For both time and accuracy, users achieved similar accuracy in less time when using heatmaps. Overall, judging from these results, we conclude that the users may take longer to finish Task 1 with similar accuracy when using \textit{Phoenixmap} . However, considering \textit{Phoenixmap} is an abstract visualization method and new to the world, this result is acceptable after the initial short training. After users better understood \textit{Phoenixmap} from Task 1, we continued to Task 2.

For Task 2 (see Fig.\ref{fig:task2}), \textit{Phoenixmap} was better on both time cost and accuracy. Users on average took 69.97 seconds to complete the task with 0.96 accuracy, while when using heatmaps they only reached 0.92 accuracy and required 76.31 seconds. The standard deviation of the time consumption difference is large; users' performance was more stable with \textit{Phoenixmap}. Remarkably, users were able to achieve significantly higher accuracy ( P-value <0.05) when using \textit{Phoenixmap} in comparison to when using heatmap.

In conclusion, we can observe that users were able to identify spatial information regarding outline and distribution density in \textit{Phoenixmap} as well as in heatmap. \textit{Phoenixmap} stably conveyed the distribution information to the users, and even significantly better than heatmap in term of accuracy. \textit{Phoenixmap} is encouraging that users can achieve such high accuracy after only one round of training within short time.

\sisetup{detect-weight,mode=text}
\renewrobustcmd{\bfseries}{\fontseries{b}\selectfont}
\renewrobustcmd{\boldmath}{}
\newrobustcmd{\B}{\bfseries}
\addtolength{\tabcolsep}{-4.1pt}

\begin{table}[h]
\centering

\begin{tabu} to 0.48\textwidth{ | c | X[c] | X[c] | X[c] | X[c] | }
\hline
\multirow{2}{*}{}  & \multicolumn{2}{c|}{TASK 1} & \multicolumn{2}{c|}{TASK 2} \\ \cline{2-5} 
                   & 8P           & 8H           & 16P          & 16H          \\ \hline
Average (Time)        & 64.06        &\B 54.46       & \B 69.97        & 76.31        \\ \hline
Average (Accuracy)    &\B 0.96         & 0.94         & \B 0.96         & 0.92         \\ \hline
STD (Time)         & \B 17.29        & 19.75        & \B 15.38        & 23.60        \\ \hline
STD (Accuracy)     &\B 0.11         & 0.13         & \B 0.08         & 0.10        \\ \hline


P-Value (Time)     & \multicolumn{2}{c|}{\B 0.002}   & \multicolumn{2}{c|}{ 0.190}   \\ \hline


P-Value (Accuracy) & \multicolumn{2}{c|}{0.401}   & \multicolumn{2}{c|}{\B 0.039}   \\ \hline
\end{tabu}
 \parbox[h]{.8\columnwidth}{\relax
P:\textit{Phoenixmap}; H: heatmap; STD: Standard Deviation
   8: 8 Graphs; 16: 16 Graphs        }
\caption{\label{tab:Stas} Wilcoxon Signed-rank Test Results for Two Matching Tasks (one-tail, paired, N=52)}
\end{table}

\subsection{Experiment \RNum{2}}

For the ranking tasks, we conducted a Spearman rank correlation to calculate the correlation between user answers and the correct order \cite{McDonald}, where the value is from -1 to 1 (-1 means totally wrong, and 1 means all correct). By this method, we can numerically examine whether two results covary. For instance, assuming the correct ranking answer is "1-2-3-4-5," "5-4-3-2-1" will return -0.999 (-1) indicating it is a completely wrong ranking answer. 

We also recorded the time consumption for each participant on each individual graph. Just as in Experiment \RNum{1}, we applied a Wilcoxon singed-rank test to compare \textit{Phoenixmap} with dotmap and heatmap visualization methods in experiment \RNum{2}.

In Table \ref{tab:Stas2}, we can see that \textit{Phoenixmap} performed significantly better than the dotmap and heatmap on ranking accuracy. For time consumption, all three visualizations are similar. From the data in Table \ref{tab:Stas2}, we can conclude that users more accurately recognized the qualitative density differences inside the spatial distribution outlines visualized by \textit{Phoenixmap}, compared to their accuracy using heatmap or dotmap. 

\sisetup{detect-weight,mode=text}
\renewrobustcmd{\bfseries}{\fontseries{b}\selectfont}
\renewrobustcmd{\boldmath}{}

\begin{table}[h]
\centering
\begin{tabu} to 0.48\textwidth{ | c | X[c] | X[c] | X[c] | }
\hline
 & Dotmap & Heatmap & \textit{Phoenixmap}\\ \hline
     
Average (Time)        &\B 16.83     &  18.02        & \B 16.90        \\ \hline
Average (Accuracy)    & 0.61               &  0.78         & \B  0.92         \\ \hline
STD (Time)         & 8.25              & \B  7.30        & 7.93       \\ \hline
STD (Accuracy)     & 0.40              &  0.41         & \B 0.24       \\ \hline
P-Value of Accuracy (Phoenix vs) & 1.13e-03 & 0.029 & - \\\hline
P-Value of Time ((Phoenix vs)) & 0.38 & 0.08 & - \\\hline
\end{tabu}

\caption{\label{tab:Stas2}Statistical Results for Ranking Task} 
\end{table}

\subsection{Experiment \RNum{3}}

From Experiment \RNum{3}, by asking each participant to evaluate 17 circles (5,6,6) in three visualizations, we collected 1088 data entries in total from 64 participants. Half of the data is for \textit{Phoenixmap} and the other half is for heatmap. For each circle, we calculated the absolute difference between the ground truth and user input. Then we computed the means and standard deviations on these numbers. The results in Table ~\ref{tab:Stas3} show that \textit{Phoenixmap}, on average, gave a 1.8-fold smaller deviation than heatmap (Means: 10.09 vs. 18.18). 

To evaluate the statistical significance of this difference, we applied a Mann-Whitney U test to examine whether two distributions are the same under the assumption of independent, ordinal responses. Strikingly, the test gave a very small p-value of 2.5e-24 (<0.01), indicating  that \textit{Phoenixmap} helps users make better density estimates, i.e., estimates with smaller deviations, than a heatmap does. In other words, users can estimate more accurately using \textit{Phoenixmap} than heatmap, at a 99 percent confidence level.

From the results, we also found that users were able to recognize and estimate the density of central regions within the tolerance level. Those areas are supposed to be the most difficult circles (e.g. circle C in the top visualizations, and circle B and F in the bottom visualizations of Fig.~\ref{fig:quant}) to estimate by \textit{Phoenixmap} since the user needs to reference segments on both sides of the circle. Compared to the overall sample test results in Table ~\ref{tab:Stas3} (Rows 1 and 2), results for the the circles in the center of the distribution (Rows 3 and 4) show that users are still able to recognize and estimate these densities better than heatmap.

\begin{table}[h]
\centering
\begin{tabu} to 0.48\textwidth{ | c |  X[c] | X[c] | }
\hline
 &  \textit{Phoenixmap} & heatmap\\ \hline
 
Mean & \B 10.09    & 18.18   \\ \hline

STD & 3.82   & \B3.27   \\ \hline

Central Mean & \B 5.22 & 7.34 \\ \hline

Central STD & \B 3.31 & 6.32 \\ \hline
     
P-Value & \multicolumn{2}{|c|}{\B 1.33e-05} \\ \hline

\end{tabu}

\caption{\label{tab:Stas3} Results for Quantitative Perceptual Estimation Task (one-tail Mann-Whitney U-Test, unpaired, N=32 each)} 
\end{table}

\subsection{User Feedback }

After these experiments, we conducted a survey with all participants. There were 32 participants submitting the survey eventually. 

Q1: What strategy did you use when reading \textit{Phoenixmap}?

We then asked the participants to rate \textit{Phoenixmap} using a Likert scale from 1 to 5 (1. Strongly disagree 2. Disagree 3. Neither agree nor disagree 4. Agree 5. Strongly agree) on the following questions. Participants were also asked to comment on \textit{Phoenixmap} after each question.

Q2: \textit{Phoenixmap} is simple to learn.

Q3: \textit{Phoenixmap} is an effective way to explore spatial distribution data.


Q4: I would like to use \textit{Phoenixmap} for future spatial distribution data visualization.

\begin{figure}[h]
     \centering
    \includegraphics[width=1\linewidth]{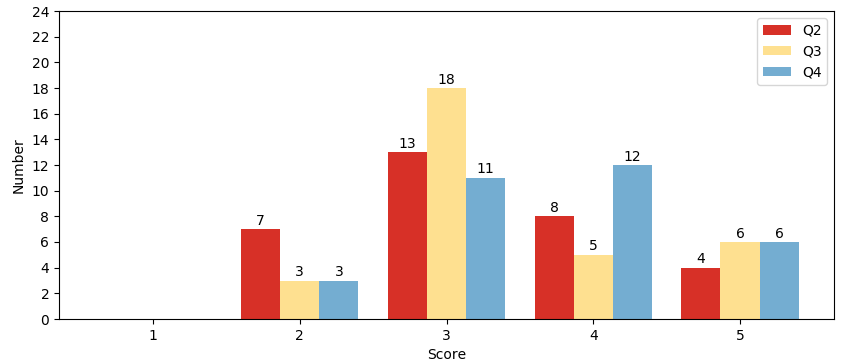}
  \caption{\label{fig:qual} Histogram for Question 2,3,4 in Qualitative Survey}
  \end{figure}

When asked about their strategies for matching the \textit{Phoenixmap} with the original point distribution, over 92\% of the participants who participated in the first two experiments compared point density areas using the corresponding segments on the outline. When working on Experiment I, the majority of participants first tried to match the outlines of two visualizations, then tried to match internal distribution by comparing segment widths with colors or points. When working on Experiment II, they simply looked for the corresponding segments and then compared the width of the segments. Most participants agreed that after the training session and experiments, they considered \textit{Phoenixmap} to be simple to use. They could easily identify the corresponding point distribution and clearly point out the normal direction and the corresponding area represented by a segment.  When the point distribution in the compared areas looked similar, the majority of participants said they needed to count the data points inside the region on the dotmaps to rank the order. However, it is a lot easier to compare the outline width in the \textit{Phoenixmap} or to compare the area colors in the heatmap. A lot of participants made mistakes when ranking using heatmaps because they ranked the order based on the most dominant color in each area. When one color within the area dominated, some participants didn't pay attention to the lighter shades in the area and tended to rank the density incorrectly.

On the Likert scale questions, the average score for question Q2 was 3.28. The participants scored 3.44 for Q3. The average score for Q4 was 3.66. The histogram in Fig. \ref{fig:qual} shows the scores distributions for each question.

Based on these results, we can conclude that the \textit{Phoenixmap} presents at least equally well as the heatmap when visualizing spatial data with respect to distribution and range. However, \textit{Phoenixmap} is more concise and allows users to develop a design by overlapping multiple ones. These improvements can supply information visualization (InfoVis) designers with more flexibility and possibility when dealing with spatial data. 

\section{Applications}
 \textit{Phoenixmap} can be applied to solve many real-world problems, such as visualizing spatio-temporal distributions of bird migration, demographic changes, or epidemiology spread\cite{eBird,spatialdata}. With \textit{Phoenixmap}'s simplicity, users can easily make clear comparisons among multiple spatial distributions according to the time changes or the differences of the objects being investigated . Fig.\ref{fig:teaser} shows a typical example of visualizing bird migration. 

\subsection{Multiple Distributions}
For further verification, we applied \textit{Phoenixmap} to more complex datasets. We utilized the public datasets provided by the IEEE Vast Challenge 2018 \cite{Phoenixmap} to visualize bird records from observations in a nature preserve over several decades. There are 19 bird types and over 2000 samples. Fig.~\ref{fig:teaser} shows the changes in six individual birds' yearly distributions visualized in \textit{Phoenixmap}. The graphs were computed from the most recent seven years' data and overlaid together in one visualization. One can easily observe the changes for a specific bird type. An online demo is prepared for readers to experience visualizations in \textit{Phoenixmap} (https://va.tech.purdue.edu/phoenixMap/). 

\begin{figure}[h]

   \begin{subfigure}[b]{0.45\textwidth}
     \centering
    \includegraphics[width=.9\linewidth]{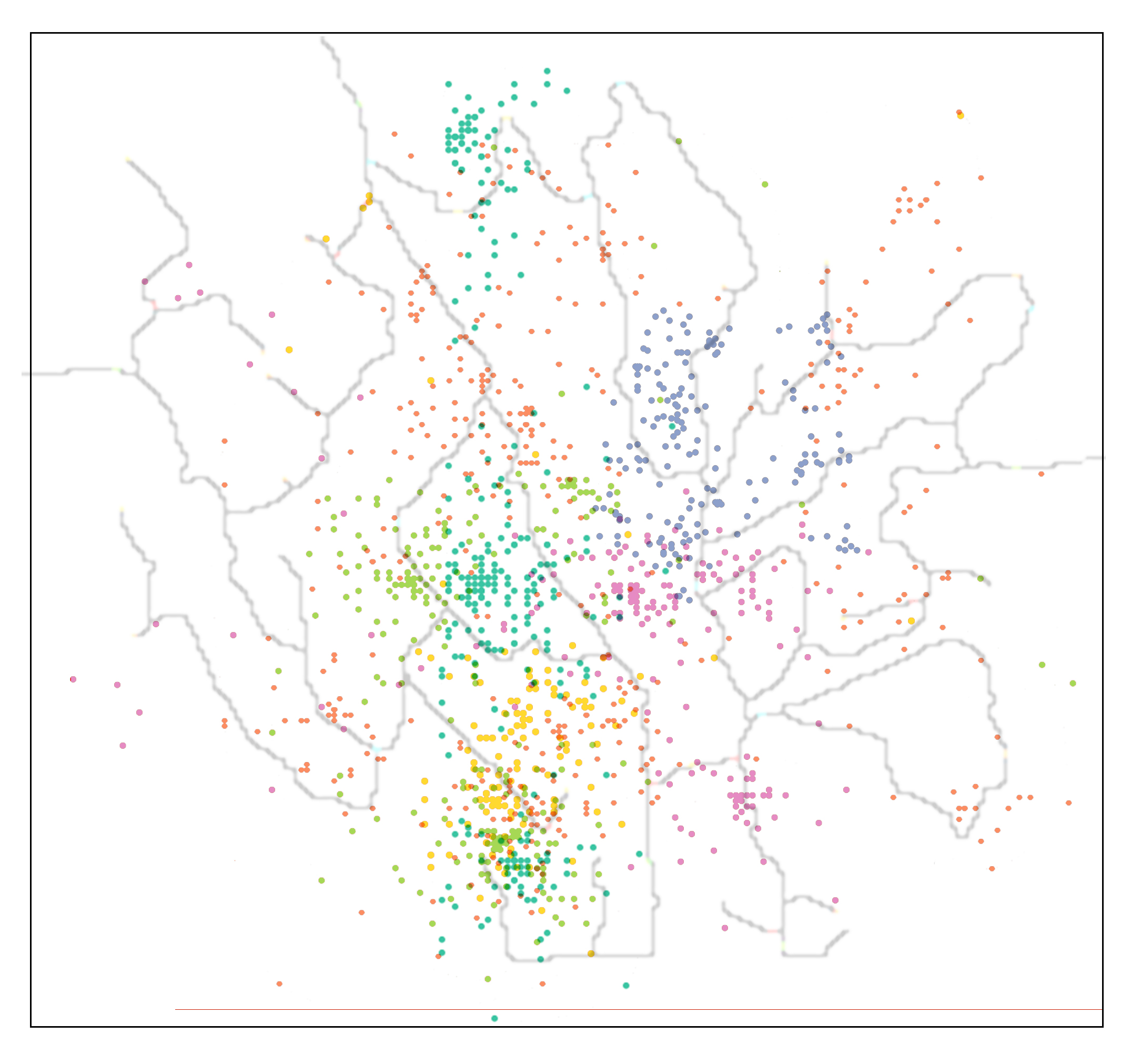}
           
  \caption{\label{fig:overlapdot} Overlapping Dot Density Maps for Multi-spatial Distribution Visualization}
  
  \end{subfigure}

\begin{subfigure}[b]{0.45\textwidth}
   \centering
  \includegraphics[width=.9\linewidth]{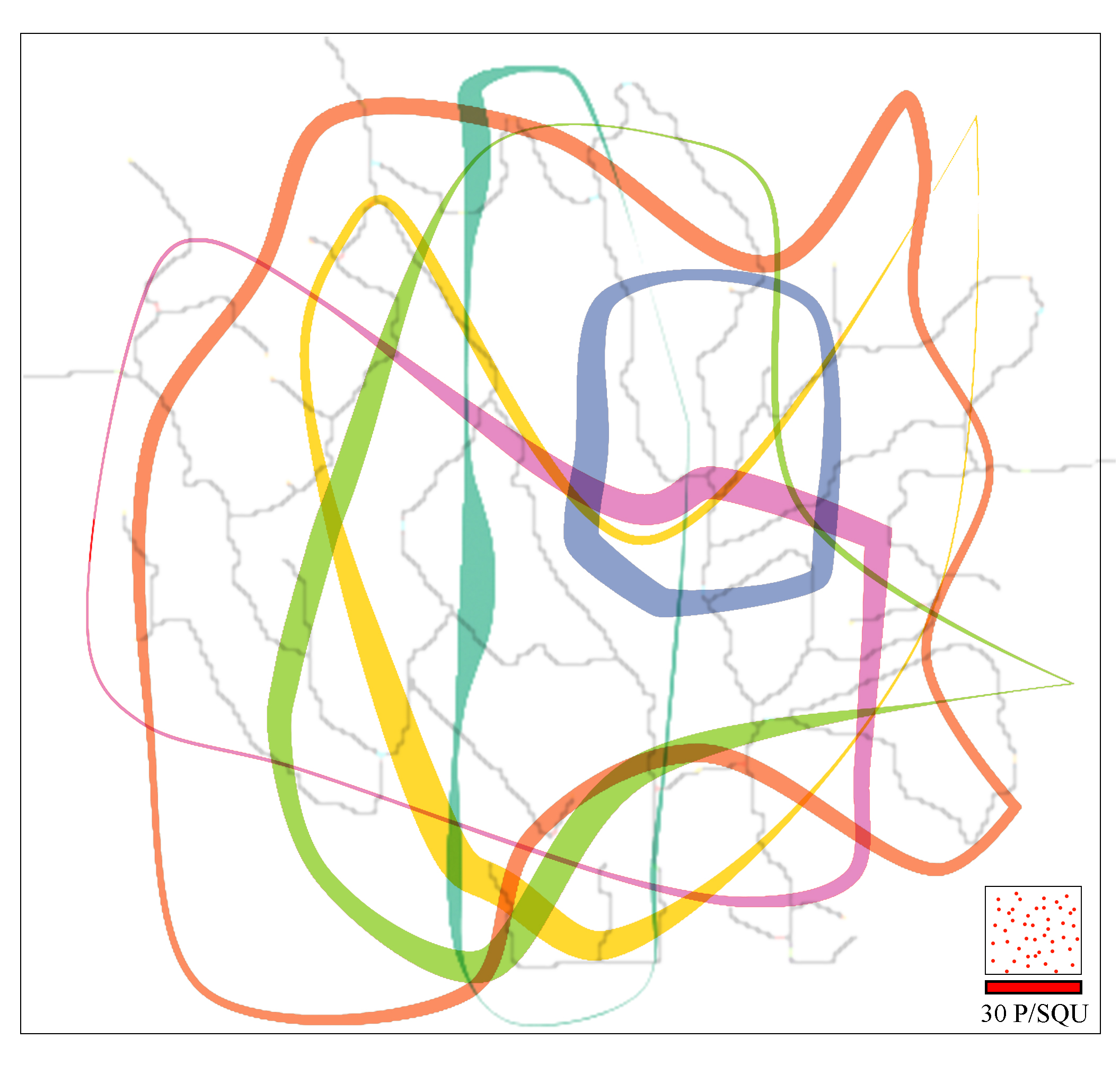}
  \caption{\label{fig:overlapgraph} Overlapping \textit{Phoenixmap}s for Multi-spatial Distribution Visualization}
  \end{subfigure}
    \caption{\label{fig:sixcolor} Active Regions of Six Kinds of Birds}
\end{figure}

Fig.~\ref{fig:sixcolor} visualizes six birds' distributions. To distinguish birds, we carefully chose contrasting colors as suggested by colorbrewer \cite{brewer2002evaluation} to represent the different types of birds. When we plotted the data for Fig.~\ref{fig:sixcolor} (a), the dotmap, users took longer time to understand the distribution information regarding densities and range outlines. From (a), some users considered themselves able to understand the distribution simply because they could see the large aggregations for each color (bird type). However, once they started exploring the distribution via \textit{Phoenixmap}, they admitted that \textit{Phoenixmap} visualizations did help them more easily identify the differences. The clear and isolated outlines allow users to determine the different types of the birds intuitively. After referring to the legend and comparing the widths (thicknesses) in \textit{Phoenixmap}, they could estimate the density distributions separately, even though there are six graphs presented together. 

In the survey described in section 5.4, we  presented Fig.~\ref{fig:sixcolor} (b) to examine participants' acceptance of \textit{Phoenixmap}. On a scale of 1 to 5, with 1 being the most negative and 5 being the most positive attitude toward the graph, the participants gave an average score of 4.2. The feedback shows that when InfoVis analysts deal with certain types of spatio-temporal data (e.g. bird migration data), \textit{Phoenixmap} allows them to analyze both space and time simultaneously in a single view by stacking multiple boundaries together on the base map. Each curved shape represents a spatial distribution over a certain period of time.

\subsection{Fixed Boundary Visualization: Off-Screen and Other Applications}
As we have stated, \textit{Phoenixmap} adopts a predetermined boundary instead of a computed clustered boundary. In some cases, the boundaries are fixed and regarded as essential clues, which can not be altered. The examples range from the walls dividing rooms to borders distinguishing political regions (e.g. counties, states, or countries). This scenario was simulated in Section 4.1.1 training task (Fig. \ref{fig:rectcircle}), where we showed the density distribution using \textit{Phoenixmap} along with the fixed boundaries. \textit{Phoenixmap} enables us to use the central space of the display for other critical information. For example, we can show people's activities within a room and display interior features such as furniture layouts in a single static visualization. Similarly, we may visualize residents' political leanings in a certain region. In real world contexts, people's political preferences are not confined by geographic boundaries, and people with different political leanings are randomly distributed within the region. In that case, we can use political boundaries as the base and overlap people's political preferences represented by colored dots on top of it. The visual result will be similar to Fig.\ref{fig:sameB}, with two or more outlines representing the distribution of people's varied political views. For these visualization applications, the main advantage of \textit{Phoenixmap} over some conventional methods is that \textit{Phoenixmap} only utilizes a minimal area on the edge, thereby creating possibilities to display other related data in the empty central area.

Similarly, if we use the edges of a display screen as the boundary and project the objects out of area on the boundary, \textit{Phoenixmap} may also be used for off-screen visualization. Off-screen visualization utilizes the areas alongside the screen edges to show  objects outside of the screen \cite{DBLP:journals/corr/JackleFR17,BaudischPatrick2003Hatf}. In a conventional \textit{Phoenixmap}, a segment's width is essentially  the density of the inside objects projected to this segment. Utilizing a projection method from \cite{DBLP:journals/corr/JackleFR17,BaudischPatrick2003Hatf}, we can visualize the number of objects outside the outline. When compared to the work of J{\"a}ckle et.al \cite{DBLP:journals/corr/JackleFR17}, \textit{Phoenixmap} lacks the capability of showing the topology and distances of objects from the outline. It is able to show the general distribution and quantity of objects outside. 

Furthermore, if we integrated both the inside and outside projections, \textit{Phoenixmap} would allow the users to make comparisons between the inside and outside objects. Considering the approach used in Fig. \ref{fig:discrete}. The black line in the figure indicates the outline of the cluster and divides the visualization area to inner side and outer side symmetrically. Alternatively, users can compute the objects inside the outline and draw the visualization using the area of the inner side of outline. Also, they can utilize the outer area of the outline to compute the visualization of the outside objects. With such an enhanced approach, users can use \textit{Phoenixmap} to visualize both sides' information and make direct comparisons between the selected and unselected samples at the same time. 

\section{Discussion}

 \textit{Phoenixmap} provides another possibility for InfoVis designers to convey spatial distributions. As an alternative to common visualization techniques, \textit{Phoenixmap} offers a practical way of producing abstract visual overviews of spatial distribution datasets, especially when there are multiple distributions belonging to different groups.  Since \textit{Phoenixmap} visualizations are hollow, we can add more spatial variables inside of each curve to convey more spatial details. For example, an InfoVis developer can place the base map to show the geographical information, or put in a text description.
 
\subsection{Intuition} From experiments, we found that users can quickly understand the idea behind the method and easily figure out how to read the visualization. For a segment on a  outline, no matter which direction its normal vector is facing, we found that the majority of users can easily figure out the direction of the normal vector and the corresponding area of the segment. Even for circles located in the center of the region, from Experiment \RNum{3} we can still see that the users can recognize that such circles are represented by two or more segments on different sides of the outline. 
 
\subsection{Estimation Accuracy}  As the results of the study suggest, one of the strengths of \textit{Phoenixmap} is that it makes it easier for the user to estimate data density and understand spatial distribution. The efficiency of the quantitative estimation on color, length and other visual variables is widely tested, and we have discussed in aforementioned literature review. Our experiment results are consistent with the discoveries from Maciejewski's effectiveness rankings of visual encodings of information \cite{maciejewski2010visual}. Thus, compared to heatmaps, in which various colors have been utilized to display density, \textit{Phoenixmap} makes it more convenient for a user to recognize the spatial density distribution over a field, using line thickness to estimate density. For dotmaps, since all points have a radius, points may sometimes overlap with each other causing overplotting issue. Also, we know from previous observations that people conduct a visual count when there are more than four objects. Thus, when there are many points, a dotmap is actually hard to read and may lead to estimation errors. Compared to using colors to estimate numerical data in heatmap, which can be perceptually difficult to interpret, \textit{Phoenixmap} uses width (thickness) to encode the numbers. The results from Experiments II and III suggest that with proper legends introduced, users can perceive and estimate the densities of the distributions. The U test result also indicates that \textit{Phoenixmap} can serve users better than either heatmap or dotmap on this estimation task. These observations are consistent with perception research on visual encoding. Cleveland and McGill \cite{cleveland1985graphical} investigated the accuracy of ten different graphical elements used to encode numerical vales and found that humans can read lengths and widths much more accurately than they can colors and shades. 

Additionally, for computing the density of the larger regions in the middle of the distribution, two or more segments may be involved, each representing a portion of the region. The majority of users can still correctly find these segments. However, the user must visually compute the density of the area using multiple segments, which makes this algorithm less intuitive with regard to such center regions. In Experiment III, we added several circles in the center of the distribution to evaluate users' perceptual estimation accuracy on those difficult regions. From the results shown in Table \ref{tab:Stas3}, we can see that users are still capable of recognizing the quantities inside those circles. In a real case,  users could refer to the surrounding regions to estimate the difficult central part. The results are acceptable. The test suggests the consistency between \textit{Phoenixmap} and heatmap.

\subsection{Scalability}  From the perspective of data size, \textit{Phoenixmap} scales well from a map with a handful of points to one with thousands or even millions of points. With the density aggregation, \textit{Phoenixmap} can handle a large number of points with no theoretical limit. In Fig.~\ref{fig:teaser}, there are about 10 to 40 observations of each bird every year. Even with these small numbers, we can construct a meaningful visualization to estimate the range and density. The sliding window function helps to smooth out uneven bumps and valleys that are not likely to happen in the real world. 

\subsection{Perceptual scalability} Perceptual scalability has been always a concern for spatial distribution visualization. Although \textit{Phoenixmap} is capable of overlaying multiple visualizations together to represent multiple spatial distributions, as in Fig. \ref{fig:sixcolor} (b), the number of outlines that can be stacked may be limited due to the risk of visual clutter. It is similar to an index graph, in which many curves are put together to represent multiple time series data, and in which clutter could occur if there are many segments crossing each other. To reduce these crossovers, one possible solution is to reduce the in-and-out waves on the outlines generated by the concave hull algorithm, e.g., by adjusting the dividing numbers or the number of nearest neighbors considered during the hull calculation. Such in-and-out waves may introduce significant crossover, so reducing the waves should also reduce problematic crossover. In extreme cases, we may use convex hull to compute the distribution outline. Also, the smoothness of these outlines may also affect perceptual scalability. According to the Gestalt laws of continuity and closure, it is much easier to perceive smooth contours than contours with abrupt changes \cite{WareC.2004IVPf}. We could adjust step 4 in our algorithm \ref{algor1} to decrease the overall curvature. To further distinguish the overlapping outlines, we may fill in the enclosed regions using a combination of textures and colors, as guided by G6.6 in \cite{WareC.2004IVPf}. But if outlines only overlap slightly, we can safely say that the \textit{Phoenixmap} is capable of stacking many outlines together. For example, we may illustrate birds' migration using dozens of \textit{Phoenixmap} outlines on the map, with each outline showing one week of birds' distribution. Since the weekly distributions trend consistently toward one direction, it is possible to put many outlines together on a map to show the consistency of the movement of birds. In such a case, the stacked curves would remain visible and legible, providing a visualization overview that meets our criteria for ease of use and integrity. 

\begin{figure}[H]
   \centering
    \includegraphics[width=.95\linewidth]{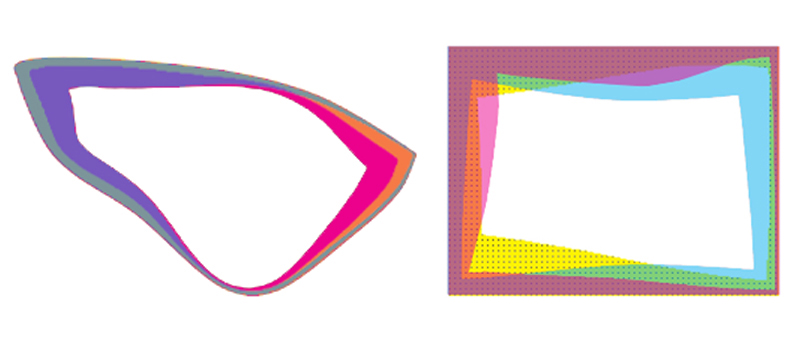}
  \parbox[t]{.9\columnwidth}{\relax
When multiple distributions share the same outline, users can adjust color or transparency, or add patterns to visualize and compare densities within the same range.}
           
  \caption{\label{fig:sameB} Multiple distributions within the same outline}
\end{figure}

In one extreme case, two distributions may have the same outline, such as in the training process for Experiment I (see Fig.~\ref{fig:rectcircle} and Fig.~\ref{fig:task1}). For such situations, we may use colors with transparency to distinguish the outlines (Fig.~\ref{fig:sameB}). This solution may only accommodate a handful of overlapped outlines before it becomes unreadable. Even though it may cause a higher cognitive and perceptional load to users, the advantage is still obvious compared to how a heatmap would perform in the same situation.

\subsection{Nesting} In certain instances, users may prefer to visualize a portion of a sample, or a "subgroup", inside a larger group. Our method could allow such users to define a boundary for the subgroup and apply the same technique in order to plot the subgroup. For instance, the blue outline in Fig.~\ref{fig:sixcolor} (b) can be imagined as a subgroup inside the larger orange outline. These subgroups' outlines can be visualized as completely nested inside the larger group's outline. With \textit{Phoenixmap}, one or many levels of subgroups can be represented coherently with the entire population. 

\subsection{Computation} The method of computing the curves requires multiple steps and contributes to the final look. The first crucial procedure is segmenting. For segmentation, the starting and ending points are not important for two major reasons. First, the segmentation number is always considerably large. In our test, we used and recommend 1000, 1500, or 3000 as empirical numbers to compute the edges. These numbers allow the algorithm to divide the original curve into segments small enough to fit in one pixel. Significantly small numbers below 100 may compute the outlines'  shapes incorrectly and loose the observability. Second, we apply sliding windows to smooth the curve's width. Therefore, the width of the curve will be continuous without significant variation.

The algorithm has some hyper-parameters to control the appearance of the final outlines. First, the number of rebuilt curve segments ($n$) and the number of divisors ($x$) determine the finely subdivided scale. From our experience, any number over 500 is enough, considering the minimum cell is one pixel. Second, the scalar $c$ is used to adjust the final widths according to the size of the graph. By adjusting these hyper-parameters, users can gain a proper scaled graph that is suitable for their dataset.

\subsection{Limitations} Despite its benefits, \textit{Phoenixmap} has limitations. Because \textit{Phoenixmap} exploits the outline to encode data density, it introduces a lack of precision as to the exact locations of each data point. However, this issue can be partially solved by implementing system interaction. We designed system interactions, e.g. mouse-over or clicking on a curve, to allow users to view the detailed points within the outline. The last graph in Fig.~\ref{fig:teaser} shows that when the user hovers over the \textit{Phoenixmap}, the dotmaps appear. 

Practically, the subjects being mapped may distribute with gaps, holes, or be separated into multiple distributions. In such cases, a simple \textit{Phoenixmap} will not be able to show these details.  Heatmap would be better for depicting these. \textit{Phoenixmap} will simply aggregate the overall density along the normal direction of a segment.

\subsection{Future Work}

There are still some open topics beyond \textit{Phoenixmap} waiting for exploration. To tackle the aforementioned limitations, some supplementary future work can be conducted. The users can implement a sophisticated clustering algorithm, e.g., Chameleon \cite{KarypisG1999Chcu}, to partition the entire dataset into multiple subgroups, and then apply \textit{Phoenixmap} to visualize each distribution with individual outlines. However, this may raise a clutter problem since it will produce more outlines. We may need to find a balance between the number of detailed clusters and the overall simplicity.  

Even though the experiment results have proven the possibility of utilizing legends to compensate for the weakness of the visualization for the central area, it demands better visualization of details. We can compute the gradients for the sub-region and encode other visual variables like single-hue gradients on the \textit{Phoenixmap}. Then, the outlines will be fairly sophisticated in terms of color and shade. It may overwhelm users' perceptions when multiple outlines are overlapped with each other. 

As stated in the aforementioned extension in 6.2, it is worth investigating the potential of using the \textit{Phoenixmap} technique to visualize the off-screen or off-bound objects in comparison with the selected objects inside the outline. How to properly calculate and visualize the distances or densities of the off-screen objects may be worth further investigating.

\section{Conclusion}

We present \textit{Phoenixmap}, an  abstract visualization technique that is suitable for visualizing multiple spatial distribution datasets. We use a closed outline with various segment widths to represent the two main attributes of spatial distribution: range and density. The user can easily recognize the boundary of a distribution and can accurately estimate the density of slices of areas adjacent to the outline segments. The outline can be a predefined fixed region or computed from an algorithm such as concave-convex hull. We divide the outline into many segments and compute the width of each segment according to the density of the facing region. The higher the density inside the range, the thicker the segment. Lastly, we smooth these segments using a sliding window for simplicity and aesthetics. Thus, a set of scattered points showing a distribution on a 2D map can be simplified and represented as a smooth curve with varying widths. In addition to the algorithm, we propose a legend design to help users understand the distribution characteristics inside the unit range. 

This technique offers several benefits: 1) The visual form is straightforward, representing spatial distribution concisely; 2) It uses considerably less screen resources and frees up the space inside the outline, which a designer can then use for detailed information, compared to conventional visualizations; 3) It is scalable by adjusting empirical number combinations, which provides adaptations and handles various object quantities; 4) It helps users make quantitative perceptual estimations; that is, when given enough legends, users can quickly estimate the density of the distribution; 5) Most importantly, \textit{Phoenixmap} enables the visualization of multiple spatial distributions by overlapping many \textit{Phoenixmaps} at the same time. 

This last characteristic of our algorithm provides further advantages over traditional geo-visualization methods such as dotmaps or heatmaps. When dealing with many different types of objects, a heatmap, by its nature, does not support overlapping, and a dotmap visualization becomes very cluttered. Our innovative solution successfully resolves this dilemma. Our empirical study shows that our algorithm is as effective and efficient as heatmap in information delivery. When representing multiple distributions simultaneously, \textit{Phoenixmap} outperforms heatmap and dotmap in terms of user comprehension. In light of the large amounts of spatial data being collected from modern technologies such as GPS, \textit{Phoenixmap} allows users to compare multiple distributions from different groups of objects, and to discern temporal changes in a distribution.

\section{Acknowledgments}

The authors would like to appreciate the support and feedback from all participants in our user study, in particular Siwei Chen from Cornell University for the contribution on the statistical evaluation.

\bibliographystyle{IEEEtran}

\vskip -3\baselineskip plus -1fil
\begin{IEEEbiographynophoto}{}
\end{IEEEbiographynophoto}
\vskip -3\baselineskip plus -1fil
\bibliography{IEEEabrv,eurovis}
\end{document}